\documentclass[twocolumn]{aastex62}
\usepackage{amsmath, amsthm, amssymb, mathrsfs}

\def \deg         {\text{$^{\circ}$}}

\begin{document}

\title{The Caltech-NRAO Stripe 82 Survey (CNSS) Paper II: On-The-Fly Mosaicing Methodology}

\correspondingauthor{K. P. Mooley}
\email{kmooley@caltech.edu}

\author{K. P. Mooley}
\altaffiliation{Jansky Fellow (NRAO, Caltech).}
\affil{NRAO, P.O. Box O, Socorro, NM 87801, USA}
\affil{Caltech, 1200 E. California Blvd. MC 249-17, Pasadena, CA 91125, USA}

\author{S. T. Myers}
\affil{NRAO, P.O. Box O, Socorro, NM 87801, USA}

\author{D. A. Frail}
\affil{NRAO, P.O. Box O, Socorro, NM 87801, USA}

\author{G. Hallinan}
\affil{Caltech, 1200 E. California Blvd. MC 249-17, Pasadena, CA 91125, USA}

\author{B. Butler}
\affil{NRAO, P.O. Box O, Socorro, NM 87801, USA}



\author{A. Kimball}
\affil{NRAO, P.O. Box O, Socorro, NM 87801, USA}

\author{K. Golap}
\affil{NRAO, P.O. Box O, Socorro, NM 87801, USA}

\begin{abstract}

Telescope slew and settle time markedly reduces the efficiency of wide-field multi-epoch surveys for sensitive interferometers with small fields of view.
The overheads can be mitigated through the use of On-the-Fly Mosaicing (OTFM), where the the antennas are driven at a non-sidereal rate and visibilities are recorded continuously.
Here we introduce the OTFM technique for the VLA, and describe its implementation for the Caltech-NRAO Stripe 82 Survey (CNSS), a dedicated 5-epoch survey for 
slow transients at S band (2--4 GHz).
We also describe the {\tt OTFSim} tool for planning dynamically-scheduled OTFM observations on the VLA, the latest imaging capabilities for OTFM in CASA, and present a 
comparison of OTFM observations with pointed observations.
Using the subset of our observations from the CNSS pilot and final surveys, we demonstrate that the wide-band and wide-field OTFM observations with the VLA can be imaged accurately, 
and that this technique offers a more efficient alternative to standard mosaicing for multi-epoch shallow surveys such as the CNSS and the VLA Sky Survey (VLASS).
We envisage that the new OTFM mode will facilitate new synoptic surveys and high-frequency mapping experiements on the VLA.

\end{abstract}


\section{Introduction}\label{sec:intro}


Survey science, both line and continuum, has become a major driver for next generation
radio facilities. A fast imaging capability with large \'entendue is a necessary requirement for all of these experiments \citep{cordes2008}. 
At centimeter wavelengths, The Australian Square Kilometer Array Pathfinder (ASKAP) and the Apertif instrument on the Westerbork Synthesis Radio Telescope (WSRT) will be using phased array technology to image large instantaneous fields of view (FoV) on the sky \citep{johnston2008,oosterloo2010}. The MeerKAT array undergoing commissioning in South Africa will achieve a high survey speed by building a large numbers of small dishes \citep{booth2012}. Traditionally, single dish telescopes have narrow fields of view but come equipped with sensitive, state-of-the-art backend receivers. In order to achieve fast survey speeds in these cases it is necessary to rapidly slew the telescope across the sky in what is called ``on-the-fly" (OTF) imaging, continually collecting data, utilizing the large instantaneous sensitivity of the telescope \citep{mangum2007}. The need for OTF is most acute in
mm/sub-mm astronomy where it has been heavily used; for example at the Nobeyama Radio
Observatory 45-m (NRO 45-m) and the Atacama Submillimeter Telescope Experiment
10-m \citep[ASTE;][]{sawada2008}, and when implementing the Legacy science programs at
the James Clark Maxwell Telescope \citep[JCMT;][]{sandell2000}. Interferometers have developed
an analogous technique called mosaicing to be used when the primary beam is smaller
than the size of the source being imaged \citep[e.g.][]{sault1996}.

In this paper we describe a new mode for the Karl G. Jansky Very Large Array \citep[VLA;][]{perley2011} that takes advantage of receiver upgrades that have enabled an order-of-magnitude increase in continuum sensitivity and we use it to implement a fast survey mode called on-the-fly mosaicing (OTFM\footnote{\url{https://science.nrao.edu/facilities/vla/docs/manuals/obsguide/modes/mosaicing}}). OTFM eliminates the slew and setup overheads and is thus ideal for either continuum or spectral line projects that require shallow, very large-area mosaics.  
In \S\ref{sec:intro:theory} we give the basic interferometric theory for OTFM.
In \S\ref{sec:otfm-vla} we describe the technical aspects relevant for the OTFM implemented for the VLA. Informed by the contents in the preceding sections, \S\ref{sec:design} gives the survey and OTFM parameters chosen for the Caltech-NRAO Stripe 82 Survey (CNSS). In \S\ref{sec:test} we describe a series of tests that were carried out for verification of the OTFM mode. 
The CNSS was designed to be a pathfinder for later wide-field surveys such as the VLA Sky Survey (VLASS\footnote{\url{https://science.nrao.edu/science/surveys/vlass}}; Lacy et al. in prep).
The scope of OTFM for a wide range of experiments on the VLA, including the CNSS and the VLASS, is given in \S\ref{sec:summary}.

\subsection{OTFM Theory}\label{sec:intro:theory}


The measurement equation of interest for an interferometric imaging problem \citep[neglecting the W term;][]{tcp99} is,
\begin{equation}\label{eq:V}
V_\nu(\vec{\bf u}_k) = \int B_\nu(\vec{\bf s}-\vec{\bf s}_p) ~ I_\nu(\vec{\bf s}) ~ e^{-i2\pi \vec{\bf u_k}(\vec{\bf s} - \vec{\bf s}_\phi)}~d\vec{\bf s}
\end{equation}

\noindent where $V_k$ is the k'th visibility, $\vec{\bf u}=(u,v)$ is the 2-D spatial frequency with respect to the phase center $\vec{\bf s}_\phi$, $\vec{\bf s}_p$ is the pointing center, 
$\vec{\bf s}=(l,m)$ is the sky coordinate, $I$ is the sky brightness (in this paper we neglect the time dependence in $I$), $B$ the primary beam pattern, and the subscript $\nu$ implies a dependence on the observing frequency.
The slew of the antennas during OTFM requires the pointing center to change with respect to the phase center with time, and the phase center is also usually updated at regular intervals. 
Hence, we generally have $\vec{\bf s}_p=\vec{\bf s}_p(t)$, $\vec{\bf s}_\phi=\vec{\bf s}_\phi(t)$.   
Within a certain integration time $\Delta t$ in which the phase center is constant while the pointing center changes, the equation should be rewritten as,

\begin{widetext}
\begin{eqnarray}
V_\nu(\hat{\vec{u}}_k) = \int  \left[\frac{1}{\Delta t}~ \int_{t_0-\Delta t/2}^{t_0+\Delta t/2} B_\nu(\vec{s}-\vec{s}_p)~dt\right]   ~ I_\nu(\vec{s}) ~ e^{-i2\pi \hat{\vec{u}}_k (\vec{s} - \vec{s}_{\phi,0})}~d\vec{s}\label{eqn:Vphi}\\
\hat{\vec{u}}_k = \frac{1}{\Delta t} \int_{t_0-\Delta t/2}^{t_0+\Delta t/2} \vec{u}_k(t) ~dt ~,~~ \hat{\vec{s}}_p = \frac{1}{\Delta t} \int_{t_0-\Delta t/2}^{t_0+\Delta t/2} \vec{s}_p(t) ~dt 
\end{eqnarray}
\end{widetext}
\noindent where $t_0$ is the time of measurement (typically taken to be the time when the pointing center coincides with the phase center) and $\vec{s}_{\phi,0}=\vec{s}_\phi(t_0)$. The time-averaged quantity $\hat{\vec{s}}_p$ is defined here for completeness, and will be used below.
The ``smearing'' described by equation~\ref{eqn:Vphi}, which occurs over time\footnote{There is another source of smearing, the baseline-based time smearing, that occurs due to $\vec{u}$ changing over the integration time, but the degree of smearing in this case is the same as that in any standard pointed observing mode.} $\Delta t$, does not introduce any phase errors (it is purely an amplitude error) as long as the time interval is sufficiently small (see below).
Furthermore, unless there are variations in the scanning (which are equivalent to pointing errors) from one antenna to another, this is purely an overall amplitude error and not antenna or baseline based.
The motion of the primary beam across a finite region on the sky during the integration time therefore necessitates the introduction of an ``effective primary beam'' (B$_{\rm eff}$).
We can rewrite equation \ref{eqn:Vphi} in the following manner using  B$_{\rm eff}$.
\begin{equation}
B_{\nu, \rm eff}(\vec{s} - \hat{\vec{s}}_p;\Delta t) = \frac{1}{\Delta t}~ \int_{-\Delta t/2}^{+\Delta t/2}  ~B_\nu[\vec{s}-\vec{s}_p(t)]~dt
\end{equation}

The accuracy of the flux densities in the image is affected by the use of the antenna primary beam instead of the effective primary beam for primary beam correction\footnote{For the purposes of the CNSS survey (specifically, the accuracy of source flux densities needed for the variability and transient search), we will show later that the use of antenna primary beam, which is predefined in standard radio astronomy software such as CASA, for primary beam correction is sufficient.}.
The associated fractional error is $B_{\nu, \rm eff}/B_\nu$, and is a function of the position within the image.
For example, moving a fifth of the primary beam (assumed to be Gaussian\footnote{In this work, we have used a simplification, that the primary beam is Gaussian, for analytical derivations (Eq.~\ref{eq:OTFSmearBeam}). The FWHM from \cite{perley2016} is used. In reality, however, the azimuthally-averaged beam within the first null differs from Gaussian by a maximum of 7\% \citep[see][this will change the smeared beam estimates by 1\% or less]{perley2016}, and additionally the beam is not azimuthally symmetric \citep[see ][]{jagannathan2017}. Further amplitude errors may also arise from the leakage of Stokes I flux into Stokes Q and U, as detailed in \cite{jagannathan2017}.}) within the data dump time gives a $<$2\% loss in accuracy, as shown below.


\section{VLA Implementation of OTFM}\label{sec:otfm-vla}

The VLA implementation of OTFM scanning employs uniform slew rates in Right Ascension (RA, $\alpha$) and Declination (Dec, $\delta$), with the coordinates of the start and end points and effective duration 
determining the rates in RA and Dec. 
A single OTFM scan (``stripe'') must be either in constant RA or constant Dec to enforce uniform dwell time on a given point in the sky.
Furthermore, uniform sensitivity requires the same separation between adjacent stripes (``rows''), $\theta_{\rm row}$, which in turn means that scanning should be in RA at constant 
Dec\footnote{OTFM on the VLA is not restricted to constant Dec stripes, however.}.
For scan in RA with rows separated by $\theta_{\rm row}$ in Dec at a fixed on-the-sky scan rate $\dot{\theta}$, the effective RA rate will be given 
by $\dot{\theta}_\alpha = \dot{\theta}/cos(\delta)$.
In the VLA Observation Preparation Tool (OPT\footnote{\url{https://science.nrao.edu/facilities/vla/docs/manuals/opt/otf}}), this is effected by specifying the 
number of phase center steps ($n_{\rm step}$) and integrations per step ($n_{\rm d}$) such that the scanning time is\footnote{In reality, 
the VLA employs a ``running start'' in order to reach the required 
constant OTFM slew speed, $\dot{\theta}_\alpha$, at the starting coordinates of the OTFM stripe. 
Hence there is an additional time $t_{\rm d} n_{\rm d}$ required at the beginning of each stripe.} 
$t_{\rm d} n_{\rm d} n_{\rm step} = \dot{\theta}_\alpha~\Delta\alpha$, 
where $t_{\rm d}$ is the data dump time (also called integration time) and $\Delta\alpha$ is the span of the stripe in RA.
The VLA data corresponding to a unique phase center, and having a duration of $t_{\rm d} n_{\rm d}$, 
is referred to as a ``subscan'', and each phase center step produces a new subscan.
In this section we describe the technical issues and formulae relevant for OTFM implemented for the VLA.


\subsection{OTFM Uniformity and Row Separation}\label{sec:otfm-vla:uniformity}

The row separation $\theta_{\rm row}$ is set by some allowed limit to the non-uniformity of the effective noise level in the resulting mosaic.
Reconstructing a mosaic at a sky position of interest is carried out by a linear weighted sum of the radio data either in the image or uv plane. 
This is equivalent to the sum,

\begin{equation}
S_\nu = \frac{1}{Z_\nu} \sum_i \frac{S_{\nu,i}~w_{\nu,i}}{b_{\nu,i}}\\
\end{equation}

\noindent where $Z_\nu=\sum_i w_{\nu,i}$ is the sum of weights (the subscript $\nu$ denotes frequency dependence), $S_i$ are the flux density values from the mosaic image (or gridded uv data) 
and $b_i=B_{\nu, {\rm eff}}(x_i,y_i;t_d)$ denotes the primary beam correction at offset $(x_i,y_i)$ from the sky position of interest.
For optimal weighting, $w_i = (b_i/\sigma_i)^2$, and using standard propagation of uncertainty, the rms in the weighted average $S_\nu$ is,

\begin{equation}
\sigma_{S,\nu}^2 = \frac{1}{Z_\nu^2} \sum_i \frac{b_{\nu,i}^2}{\sigma_{\nu,i}^2} = \frac{1}{Z_\nu}
\label{eqn:uncertainty_weighted}
\end{equation}

\begin{figure}
\centering
\includegraphics[width=3.2in]{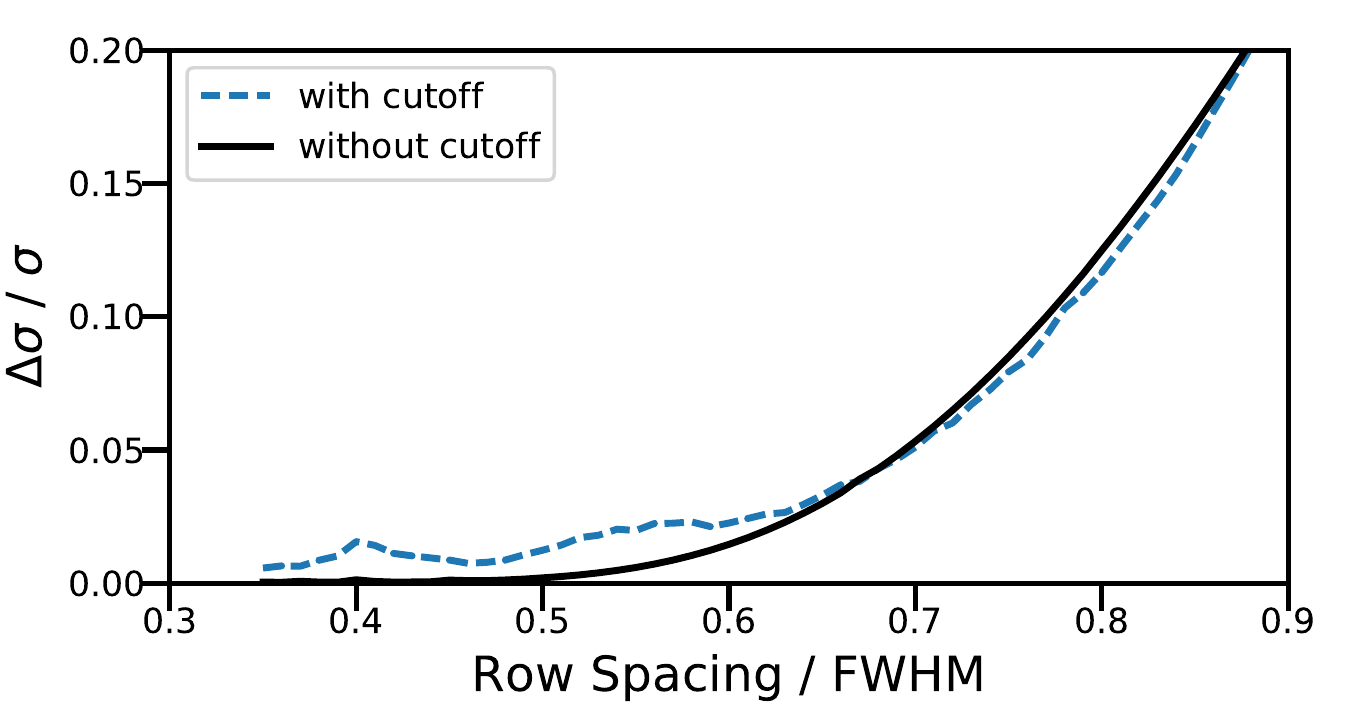}
\caption{Uniformity of linear mosaic as a function of the OTFM row spacing.
The uniformity is characterized by the maximum normalized deviation ($\Delta\sigma/\sigma$).
The black solid curve corresponds to the ideal case in which there is no beam response cutoff for the scans included in the mosaic. 
The blue dashed curve is for the case where the cutoff is where the Gaussian beam falls to 17\% (0.8$\theta_{\rm pb}$).
Around $\theta_{\rm row}/\theta_{\rm pb}\simeq$0.55--0.65, the blue curve plateaus with a non-uniformity of around 2\%, which is 
the best that one can achieve using normal imaging procedures.}
\label{fig:OTFMuniformity}
\end{figure}

\noindent The mosaic RMS noise is therefore simply the inverse of the sum of mosaic weights, $Z$.
The normalized maximum deviation is defined as,
\begin{equation}
\frac{\Delta \sigma}{\sigma} = 2~\frac{\sigma_{\rm max}-\sigma_{\rm min}}{\sigma_{\rm max}+\sigma_{\rm min}}
\label{eqn:max_dev}
\end{equation}
\noindent where $\sigma_{\rm max}$ and $\sigma_{\rm min}$ are the maximum and minimum RMS noise in the reconstructed mosaic assuming uniform intrinsic noise per scan.
We simulated a grid of pointings with uniform RMS noise and varied the row separation. 
For each value of row separation, we numerically calculated the effective mosaic RMS noise given by equation~\ref{eqn:uncertainty_weighted} and then the normalized maximum deviation 
as per equation~\ref{eqn:max_dev}.
The resulting curve is shown as the solid black curve in Figure~\ref{fig:OTFMuniformity}.
We repeated this exercise restricting each pointing to within the 20\% power point (0.8$\theta_{\rm pb}$).
The curve obtained using this constraint is shown as the dashed cyan curve in Figure~\ref{fig:OTFMuniformity}.
While the unrestricted curve gives a perfectly uniform RMS noise below $\theta_{\rm row}=0.8\theta_{\rm pb}$), the restricted curve plateaus with a 
non-uniformity of around 2\%, which is in practice the best achievable using standard imaging procedures.

\subsection{OTFM Smearing}\label{sec:otfm-vla:smearing}
\begin{figure}[htp]
\centering
\includegraphics[width=3.4in]{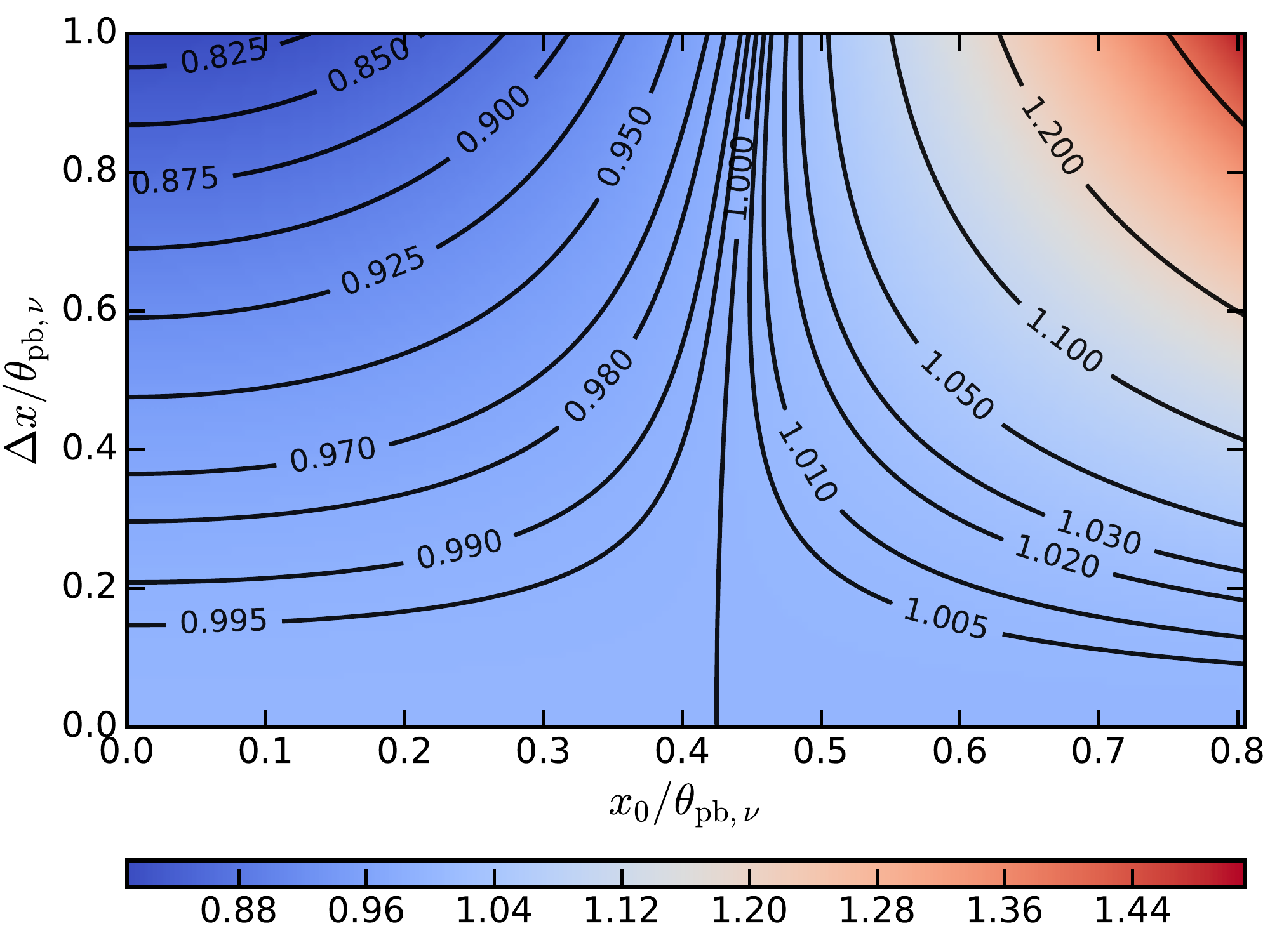}
\caption{The OTFM smeared beam at 3 GHz.
The plot shows the fractional change in the flux density with respect to the true flux density (colorbar at the bottom) when only a single, time invariant, primary beam correction is applied to each scan.
The x-axis is the distance, in units of the primary beam FWHM and in the direction of the OTFM scan (in this case, the right ascension), from the phase center in the image plane.
The y-axis is the slew of the antennas, in units of the primary beam FWHM, within each subscan. 
For the CNSS, $\Delta x/\theta_{{\rm pb},\nu}=0.36$ for each subscan (4s-long).
Alternatively, this plot can also be used to find the smearing within each data dump (for the CNSS, $\Delta x/\theta_{{\rm pb},\nu}=0.05$ using $t_d=0.5$s)
Note that the fractional change in the flux density does not depend on the distance in the direction orthogonal to the direction of the OTFM scan, as described in \S\ref{sec:otfm-vla:smearing}.}
\label{fig:OTFSmearBeam}
\end{figure}

We now derive an expression for the fractional change in the flux density of sources if the motion of the primary beam 
within a subscan (duration in which the phase centre is constant) or data dump time is neglected.
Describing the primary beam with a 2-D Gaussian function we have,
\begin{equation}
 B_\nu(x,y) = e^{-(x^2+y^2)/2\rho_\nu^2}
\end{equation}

where x and y are cartesian coordinates in the image plane defined with respect to the pointing center.
We assume for simplicity that the motion of the antennas is along the x-direction.
For data taken over the interval from $t_0-\Delta t/2$ to $t_0+\Delta t/2$ at OTFM scanning rate $\dot{\theta}$ in the x-direction, 
a visibility centered at an offset point $(x_0,y_0)$ in the primary beam will see an effective beam value of:
\begin{widetext}
\begin{align}
 B_{\nu, \rm eff} (x_0,y_0;\Delta x) &= \frac{1}{\Delta x} \int_{x_0-\Delta x/2}^{x_0+\Delta x/2} B_\nu(x,y_0) ~{\rm d}x \\
            &= \frac{e^{-y_0^2/2\rho_\nu^2}}{\Delta x} \int_{x_0-\Delta x/2}^{x_0+\Delta x/2} e^{-x^2/2\rho_\nu^2} {\rm d}x \nonumber \\
            &= \frac{\rho_\nu}{\Delta x} \sqrt{\frac{\pi}{2}} e^{-y_0^2/2\rho_\nu^2} \left[ {\rm erf}\left(\frac{x_0+\Delta x/2}{\sqrt{2}\rho_\nu}\right) - {\rm erf}\left(\frac{x_0-\Delta x/2}{\sqrt{2}\rho_\nu}\right) \right] \nonumber\\
            &= \frac{\rho_\nu}{\Delta x} B_\nu(x_0,y_0) ~e^{x_0^2/2\rho_\nu^2} \sqrt{\frac{\pi}{2}} \left[ {\rm erf}\left(\frac{x_0+\Delta x/2}{\sqrt{2}\rho_\nu}\right) - {\rm erf}\left(\frac{x_0-\Delta x/2}{\sqrt{2}\rho_\nu}\right) \right] \label{eq:OTFSmearBeam}
\end{align}
\end{widetext}
where $\rho_\nu=\theta_{{\rm pb},\nu}/2.355$ is the standard deviation of the primary beam at frequency $\nu$, $\theta_{{\rm pb},\nu}$ is the FWHM of the primary beam, 
and $\Delta x=\dot{\theta}~\Delta t$ is the slew of the antennas within time interval $\Delta t$.
The net effect of the smearing is an overall change in the flux densities, and the fractional error, $B_{\nu, \rm eff}/B_\nu$, is shown in Figure~\ref{fig:OTFSmearBeam}.
We see from this figure that, for $\Delta x<0.3~\theta_{\rm pb}$, the fractional error is $<$2\%.
Also, according to this expression, the fractional error is independent of the y-coordinate (orthogonal direction to the motion of antennas).

\begin{figure}[htp]
\centering
\includegraphics[width=3.5in,viewport=100 0 580 450,clip]{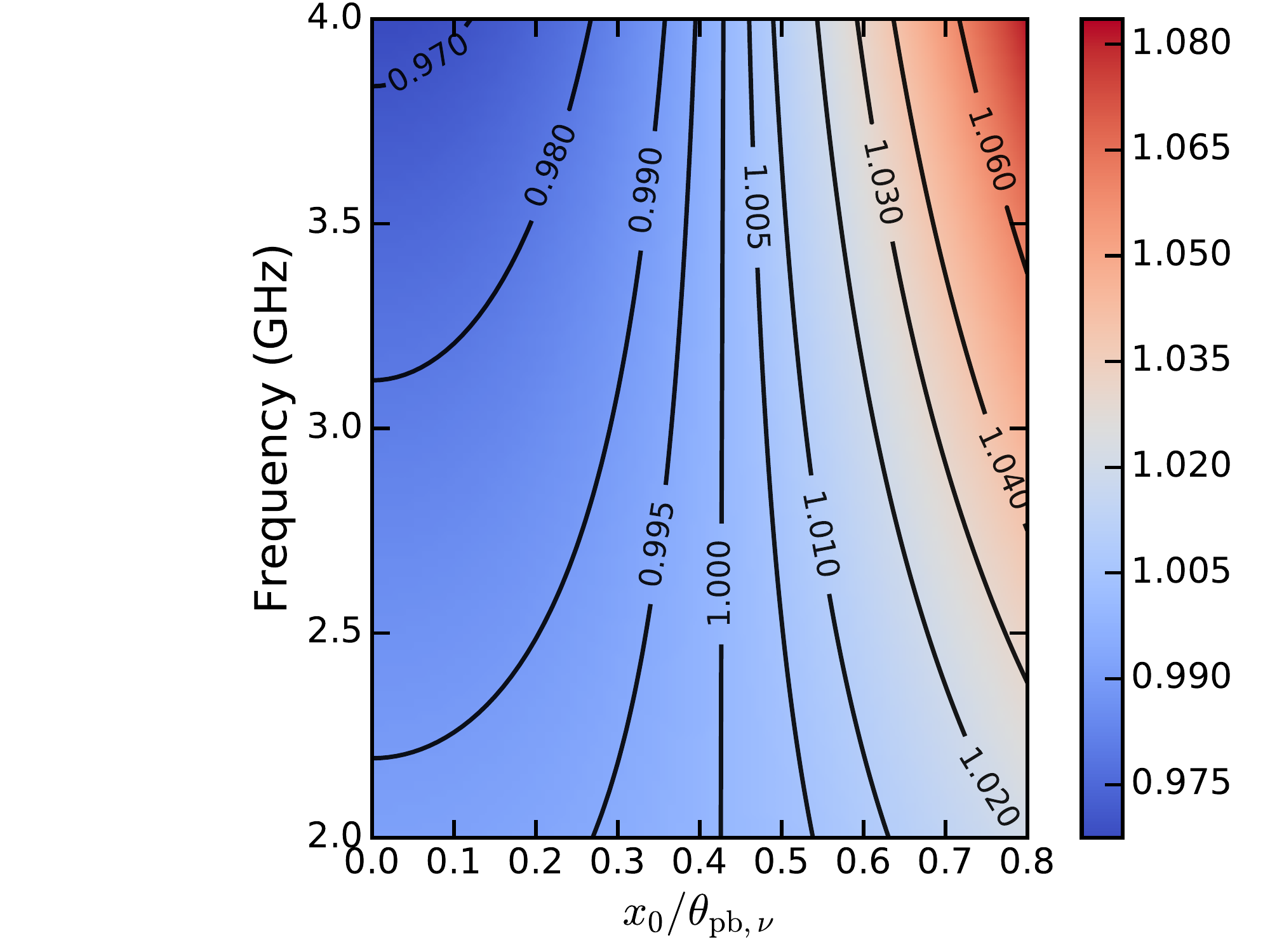}
\caption{The OTFM smeared beam plotted as a function of frequency within S band.
The plot shows the fractional change in the flux density with respect to the true flux density (colorbar at the right) when only a single, time invariant, primary beam correction (at the 
corresponding frequency shown on the y-axis) is applied to each scan.
The x-axis is the same as in Figure~\ref{fig:OTFSmearBeam}.
Parameters relevant for the CNSS ($\Delta x/\theta_{{\rm pb},\nu}=4\arcmin/(42\arcmin/\nu_{\rm GHz})=0.36(\nu/3~{\rm GHz})$ per subscan) are used.
See \S\ref{sec:otfm-vla:smearing} and \S\ref{sec:design:OTFM} for details.
}
\label{fig:OTFSmearBeam_Freq}
\end{figure}

As noted above, the OTFM smeared beam, and therefore the error introduced in the flux densities, is a function of frequency.
Since the beam gets smaller with frequency, the error is larger at the top of the band (i.e. at higher frequencies) than the bottom of the band.
Figure~\ref{fig:OTFSmearBeam_Freq} shows the frequency dependence of the smeared beam relevant for the CNSS ($\Delta x/\theta_{{\rm pb},\nu}=0.36(\nu/3~{\rm GHz})$ per subscan; see \S\ref{sec:design:OTFM}).

\subsection{Data Dump Time and OTFM Scan Rate}\label{sec:otfm-vla:dump}
To avoid phase errors\footnote{see \cite{daddario2000} for an analogous approach for estimating the coherence loss for a linear scan with a fixed phase center} 
within the data dump time $t_{\rm d}$, it is necessary to have $\vec{s}_p(t)$ varying by a magnitude less than 
the distance associated with the tolerable aliasing \citep[$d_{\rm alias}$; see][]{pety2001,pety2010} within this time interval.
It requires the following condition to be met \citep{pety2010}.
\begin{equation}
t_{\rm d} \ll \frac{d_{\rm alias}}{d_{\rm max} \omega_{\rm earth}} ~\equiv ~ t_{\rm d} \ll \frac{6900 s}{\theta_{\rm alias} / \theta_{\rm syn}}
\label{eq:td}
\end{equation}

\noindent where $d_{\rm max}$ is the maximum baseline length, $\omega_{\rm earth}$ is the angular velocity of a spatial frequency due to the Earth rotation ($7.3\times$10$^{-5}$ rad s$^{-1}$). 
$\theta_{\rm alias}$ and $\theta_{\rm syn}$ are the angular values corresponding to the field of view giving a tolerable aliasing, and the synthesized beam respectively.

Conversely, the lower limit to the dump time (and equivalently an upper limit on the OTFM scan rate) is set by the data rate, $R$. 
For the VLA, $R=45 \times n_{\rm spw} n_{\rm chan} n_{\rm pol}/(16384\times t_d)$.
For the regular S band setup, the number of spectral windows is $n_{\rm spw}=16$, channels is $n_{\rm chan}=64$ (2 MHz-wide, over a 2 GHz bandwidth), 
polarizations is $n_{\rm pol}=4$.
This sets the lower limit of $t_d=0.45$ seconds, assuming standard data rates (maximum rate of 25 MB/s).


\subsection{Image Mosaic Construction}\label{sec:otfm-vla:imaging}

Standard imaging techniques are applicable to OTFM data, and the dirty image is the fourier transform of visibilities described in Eq.~\ref{eqn:Vphi}.
The image mosaic for OTFM observations can be constructed through joint mosaicing or linear mosaicing of all scans within a region of interest.
In joint mosaicing, the uv data of all the scans of interest are gridded together using the aperture function and then deconvolved.
In linear mosaicing, each scan is imaged separately and later combined using primary beam weights.

Either the smeared beam ($B_{\nu,{\rm eff}}$) or the actual beam ($B_{\nu}$) can be used for primary beam correction.
Some amplitude error will be effected\footnote{The thermal rms noise continues to follow the radiometer equation, of course.} 
by the use of either of these beams, as described in \S\ref{sec:otfm-vla:smearing}, unless the data within each dump time is corrected with the appropriate smeared beam.
Note that there are additional sources of amplitude error, resulting from antenna pointing error and error in the absolute flux scale calibration, 
amounting to $\sim$5\% \citep[e.g.][]{thyagarajan2011,mooley2013}.
Taken in quadrature together with the OTFM smearing error, the amplitude errors should be $\sim$5--15\% for OTFM observations (the maximum error occurring due to the use of the actual beam rather than the smeared beam; assuming $\Delta x/\theta_{{\rm pb},\nu}<0.5$), 
with the most severe errors seen in sources that are well beyond the half power point of the primary beam throughout the OTFM scans. 


%
Unless the motion of the primary beam within the pointing (or scan) is taken into account, the flux densities are added linearly without any weighting, which might limit the dynamic range.
Faster scanning rates are thus expected be more limiting in dynamic range.

Currently, the pointing direction of the antennas is recorded at regular intervals and stored as a pointing table in the Science Data Model (SDM) format of the VLA observations.
The motion of the antennas (i.e. the smeared beam) can be corrected in the imaging step with the CASA task {\it tclean} using the pointing table, as briefly described in the following subsection. 
However, the pointing table feature was implemented in the VLA OTFM observations starting 2016, and for the CNSS there are no such tables in the SDM data files.
The $\sim$5--15\% are, however, acceptable for the primary science goal of the CNSS, which is finding highly variable and transient sources.

\subsection{Joint Deconvolution and A-projection}\label{sec:intro:jd}
Image-plane mosaicing does not account for the frequency-dependence of the primary beam, thus introducing spectral (and therefore amplitude) errors. Mosiacing in the uv domain \citep[or joint deconvolution incorporating A-projection; e.g.][]{bhatnagar2008,bhatnagar2013}, where the Fourier transform of the beam of each antenna at different frequencies can be taken into account during the convolution step, is therefore preferred for wide-bandwidth OTFM data. This method is currently available in {\tt clean} (using ftmachine='mosaic') and {\tt tclean} (gridder='mosaicft') tasks within CASA (version 5.3.0). In CASA, joint deconvolution is done on the dirty mosaic image (or residual) using an approximate point-spread function (PSF; using one for the OTFM scans near the center of the image). Due to the approximate PSF used for all the scans, many Cotton-Schwab major cycles are carried out to correct for the errors from the minor cycle deconvolution stage. Currently, the primary beam correction (using the actual beam, not the smeared beam) is done for at every dump time for each baseline. Some amplitude errors will therefore be introduced if the beam is moving a large fraction within the dump time. A detailed description of imaging with the new CASA task, {\tt tclean}, and its application to OTFM data will be given elsewhere (Rau et al. in prep).



\subsection{Effective Survey Speed}\label{sec:intro:SS}

For a large OTF region covered, the raw survey speed is given by ${\rm SS}=\Omega_{\rm pb}/\tau$, where $\Omega_{\rm pb}=0.566\theta_{\rm pb}^2$ is the area under the primary 
beam and $\tau$ is the on-source time required for reaching the desired point-source sensitivity, $\sigma$.
Since CNSS is a broadband mosaicked survey, the effective (weighted arithmetic mean) frequency of the images, $\bar{\nu}\simeq2.8$\,GHz, is different from the nominal 
observing (arithmetic mean) frequency, $\nu_{\rm obs}\simeq3.0$\,GHz \citep{condon2015}.
Here we have ignored the fact that some spectral windows at the bottom of the band are lost due to satellite-induced RFI (see \S\ref{sec:img:RFI}).
The effective area under the beam therefore becomes $\bar{\Omega}_B=\Omega_B(\bar{\nu})=0.566\bar{\theta}_{\rm pb}^2$.
The VLA primary beam (FWHM) as a function of frequency is approximately $\theta_{{\rm pb},\nu}\simeq(42\arcmin/\nu)$ \citep{perley2016}.
Thus, $\bar{\Omega}_B=0.035$ deg$^2$.
Now, the effective weighted system equivalent flux density (SEFD) over the VLA S band is 350.7 Jy (E. Momjian, private communication).
Therefore,

\begin{equation}
\label{eqn:sefdtime}
\tau=\frac{<\rm SEFD>^2}{n_{\rm p}\,\sigma^2\,\eta^2\,N\,(N-1)}=7.45 s\,\left(\frac{\sigma}{100~\mu{\rm Jy}}\right)^{-2}
\end{equation}

where $N=26$ is the number of antennas, $n_{\rm p}=2$ is the number of polarizations, $\eta=0.92$ is the correlator efficiency.
For S band we therefore get,

\begin{equation}
\label{eqn:SS}
{\rm SS}=16.9\,{\rm deg}^{2}\,{\rm hr}^{-1}\,(\sigma/100~\mu{\rm Jy})^{-2}
\end{equation}


%


\section{CNSS Survey Design}\label{sec:design}

\subsection{Design of the OTFM Observations}\label{sec:design:OTFM}


The CNSS \citep[project codes VLA/13B-370, VLA/15A-421, PI: G. Hallinan; see][for the pilot survey]{mooley2016} is designed to survey the entire 270 deg$^2$ of the SDSS Stripe 82 region over 5 epochs.
With the aim of characterizing the sub-mJy transient population, the required RMS noise is 80 $\mu$Jy beam$^{-1}$ (with natural weighting) per epoch.
From equation~\ref{eqn:SS} we can calculate the survey speed as 10.8 deg$^2$ hr$^{-1}$.
For ease of dynamic scheduling, we designed individual scheduling blocks (SBs) to be of 3 hours duration.
Each block was equipped with its own flux calibrator and polarization angle calibrator (3C48), polarization leakage calibrator (3C84), and nearby phase calibrators (see below).
In each hour we are able to survey 22.5 deg$^2$ (9 deg in RA and 2.5 deg in Dec) in 3 hr, including overheads.
The on-target time per SB was 2.25 hours, and overhead was therefore 33\%. 
Our coverage of Stripe 82 with OTFM was in the region $\alpha\in[-50,58]$ and $\delta\in[-1.25,1.25]$, divided into 12 SBs\footnote{Regions R1--R12, with each region spanning 9 deg in RA and 2.5 deg in Dec.}.

The desired RMS noise requires an OTFM scan rate of $\dot{\theta}_\alpha\simeq\dot{\theta}/cos(0\deg)=1\arcmin/s$.  
In order to get a fairly uniform sensitivity across the survey region, we chose a row separation of $\theta_{\rm row}=\theta_{\rm pb,2.8 GHz}/\sqrt{2}=10.6\arcmin$.
This gives a normalized maximum deviation (cf. equation~\ref{eqn:max_dev}) of 0.05.
We used 15 stripes to cover the declination range of Stripe 82.
At the scan rate chosen, each stripe took 9.25 min to get 9 min on-source integration (3\% scan overhead\footnote{For the CNSS, an OTFM scan required a ''start-up'' time of $\sim$15--20 s
at the beginning of a stripe and a ``running-stop'' overhead of 4s.}). 
Groups of two stripes were interspersed with phase calibrator scans (with a single stripe left over in the block; typically 1 min of overhead for every 20 min of OTFM). 
There were extra observations of calibrators from adjacent blocks included to help link the blocks. 
This was a fairly conservative strategy, but control of calibration errors was important for the CNSS observing program, given the proximity to the Clarke Belt (see also \S\ref{sec:img:RFI}).


In order to choose the data dump time, we refer to equation~\ref{eq:td}.
For a dynamic range of a few thousand, $\theta_{\rm alias} / \theta_{\rm pb} \simeq 5$ ~\citep[Table 4 of ][]{pety2010}. 
Hence, and further using the parameters associated with the CNSS observations at a frequency of 3 GHz ($\theta_{\rm pb}\simeq14\arcmin$, $\theta_{\rm syn}\simeq2.5\arcsec$), we arrive at,
$\delta t \ll 1400 s/(\theta_{\rm pb} / \theta_{\rm syn}) \simeq 4.2~s$.
For the CNSS survey we have used an data dump time of 0.5 s for the first three epochs and 1s for the final two epochs (due to data rate constraints; \S\ref{sec:otfm-vla:dump}).
This implies 30 dumps per primary beam (at 3 GHz; 0.5 s dumps).
The fractional amplitude error in source flux densities introduced due to the motion of the primary beam within the dump time is $<$1\%.


The correlator phase center was stepped every 4s (i.e. every 4\arcmin~or $\Delta x/\theta_{{\rm pb},\nu}=0.36$ at 3 GHz).
Each subscan was thus 4s long and had the phase center same as the pointing center at the center of the subscan.
This corresponds to $\lesssim$5\% error in the flux densities of sources (within the 20\% attenuation radius of the primary beam) due to the use of the antenna primary beam instead of the effective primary beam (see Figure~\ref{fig:OTFSmearBeam_Freq}).
Note that, for the gain calibrator observations, standard pointed mode observations are used.


\begin{figure}[htp]
\centering
\includegraphics[width=3.3in]{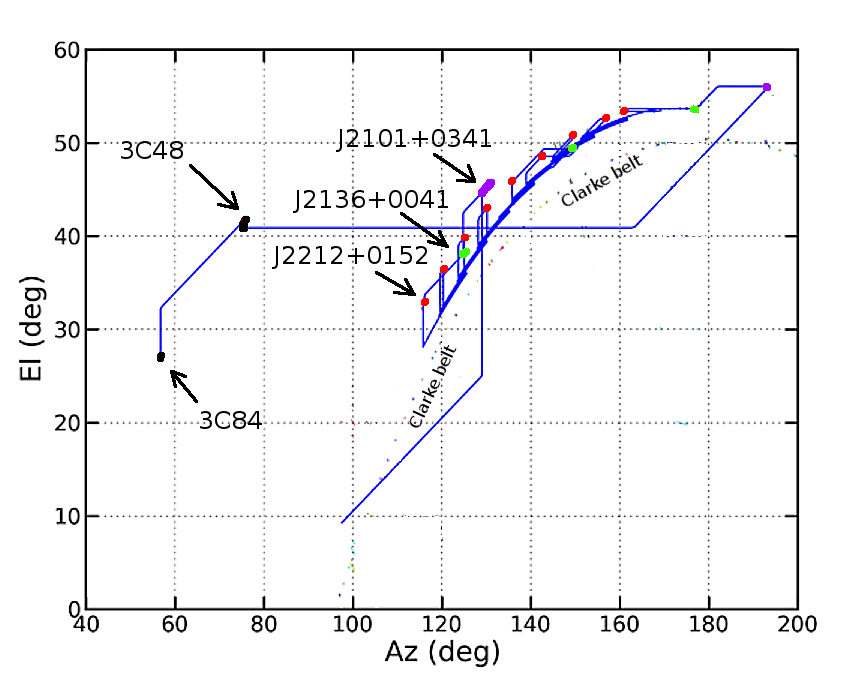}
\caption{OTFM observation planning for dynamic scheduling. 
This plot shows the azimuth (AZ) and elevation (EL) simulated for one of our observing blocks.
The assumed starting position of the antennas is 0$^o$ in right ascension and declination, corresponding to AZ$\simeq$100$^o$ and EL$\simeq$10$^o$ at the specified start LST.
The AZ and EL of the antennas through the observation is denoted by the blue line.
Tertiary and secondary calibrators are denoted by magenta and green circles respectively.
The gain calibrator source is represented by red circles, and the flux and polarization calibrators, 3C48 and 3C84, are denoted by black circles.
The known Clarke belt satellites (database maintained by Vivek Dhawan, NRAO) on the sky are shown as thin colored lines.
See \S\ref{sec:design:OTFSim} for details.}
\label{fig:plan}
\end{figure}

\subsection{Construction of the Observing Blocks via {\tt OTFSim}}\label{sec:design:OTFSim}

The CNSS is designed to be a pathfinder for future wide-field surveys with the VLA, and was therefore executed with completely-dynamic scheduling.
With the VLA dynamic scheduling, constraints can only be placed on the local sidereal time (LST); the exact date and start time of the observations cannot be predicted beforehand.
This necessitated the design of observing blocks such that each block is self-contained, with the standard calibrator observations.
Additionally, Stripe 82 being close to the Clarke belt, some azimuth ranges need to be avoided to mitigate strong RFI due to satellite downlink signals.
In order to determine the desirable LST ranges for our observations, we therefore wrote a Python-based program, 
{\tt OTFSim}\footnote{Available via GitHub.} 
to simulate the position of the antennas in azimuth and elevation.
Given a simple user input, {\tt OTFSim} computes the entire sequence of an observing block, allows the visualization of the antenna positions and optimization of block.
It outputs the source list and schedule files that can be uploaded to the Observer Preparation Tool (OPT) of the VLA.

Figure~\ref{fig:plan} shows the plot of the azimuth and elevation simulated for one of our scheduling blocks executed on 21 Dec 2013.
The assumed starting location of the antennas is right ascension and declination of 0$^o$, corresponding to az$\simeq$100$^o$ and el$\simeq$10$^o$ at the specified start LST of 18h.
The blue line denotes the motion of the antennas during the observation.
At the beginning of the observation, the antennas slew to a tertiary calibrator (denoted by magenta circles), J2101+0341, which is observed once at the beginning and once towards the end of the observation.
A secondary calibrator (denoted by green circles), J2136+0041, is then observed, followed by the gain calibrator (red circles), J2212+0152.
The survey region is then observed with OTFM interleaved with phase calibrator observations.
At the end of the scheduling block, the flux calibrator, 3C48, and the polarization calibrator, 3C84 are observed (black circles).

\subsection{RFI Monitoring}\label{sec:img:RFI}
\begin{figure}
\centering
\includegraphics[width=3.5in]{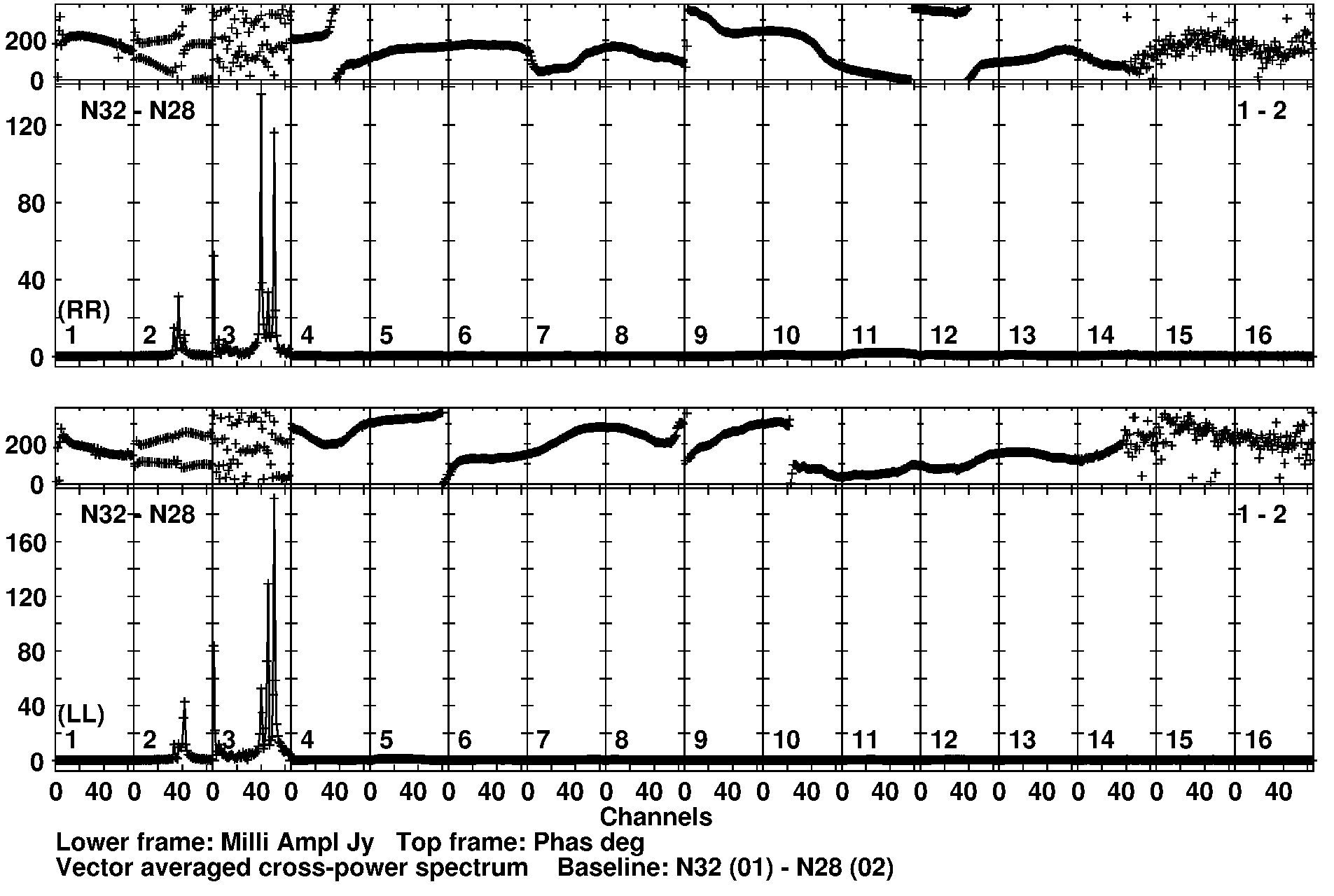}
\caption{Raw spectra (2--4 GHz) of the gain calibrator source, J2212+0152, from the 21 Dec 2013 epoch, i.e. phase vs. channel (upper frame in the top and bottom panels) 
and amplitude vs. channel (lower frame in each panel) for the RR and LL polarizations (top panels and bottom panels respectively). 
All baselines and all pointings of the gain calibrator source have been combined to produce these plots. 
Note the large-amplitude RFI in the SPWs 2 and 3, and the noisy phases in SPWs 14--16.}
\label{fig:RFI}
\end{figure}

Since Stripe 82 is close to the Clarke belt, radio observations are prone to severe RFI from satellites in geostationary and geosynchronous (GSO) orbits.
Two spectral windows (SPWs), between 2.125--2.375 GHz are severely and irreparably affected by RFI from the Satellite Digital Audio Radio Service (DARS) and satellite downlink.
The RFI in the frequency range 3.62--4.00 GHz, which is also due to satellite downlink, is low-level in amplitude, but it distorts the phase information quite significantly.
GSO satellites seen by the VLA have not been individually characterized in terms of downlink frequencies and polarizations, but the locations of some of these 
satellites are known (Figure~\ref{fig:plan}).
The RFI in the gain calibrator (J2212+0152) observations from the region R3 epoch E1 of the CNSS (observed on 2013 December 21) is shown in Figure~\ref{fig:RFI}.

\begin{figure*}[t]
\centering
\includegraphics[width=5.5in,angle=-90]{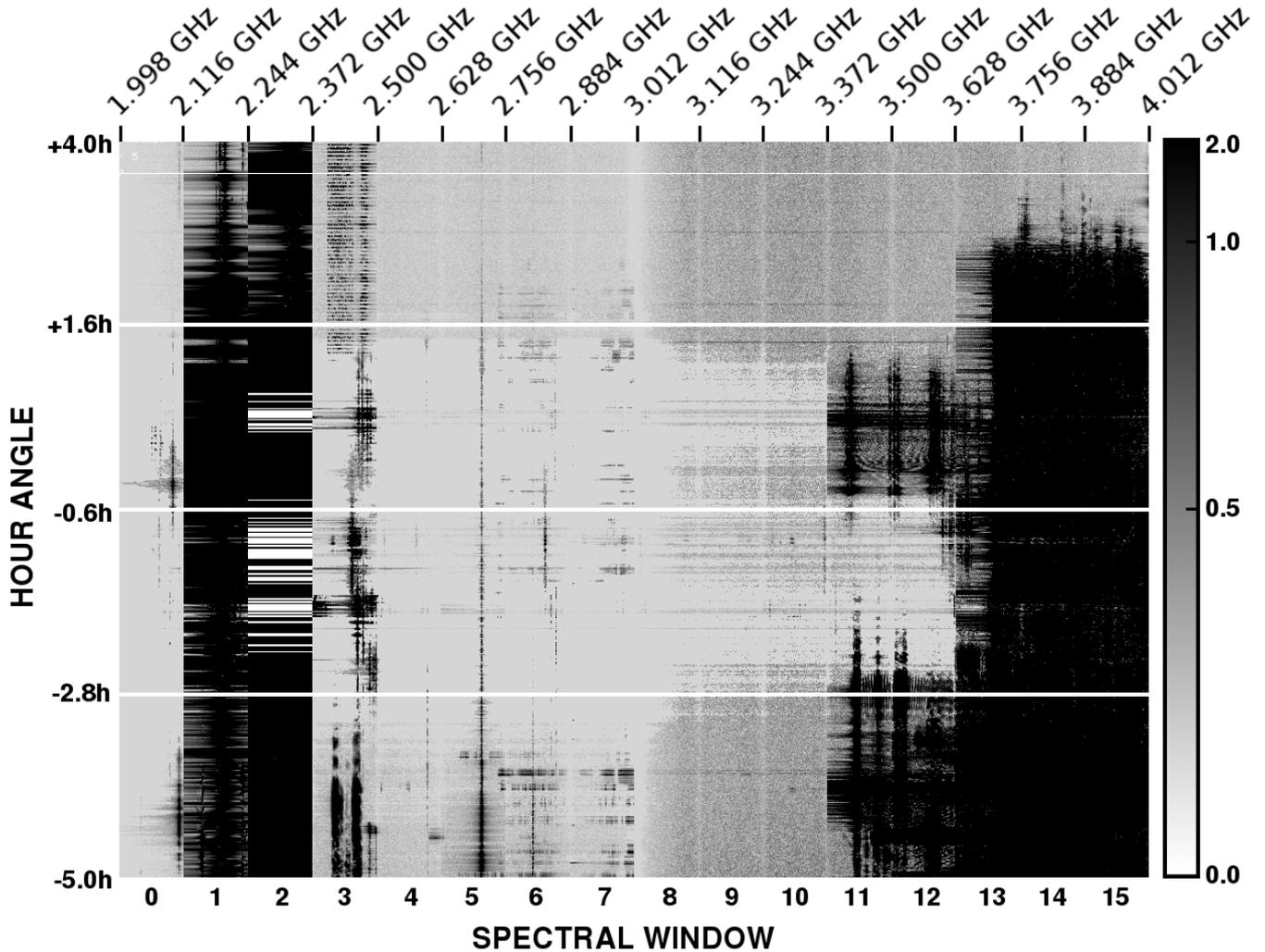}
\caption{A dynamic spectrum (in LL polarization) on a single baseline. The frequency axis runs from 2 to 4 GHz. 
The color scale runs from 0 Jy (white) to 2 Jy (black).
The plot demonstrates that the SPWs 1--2 and 13--15 (indexing different from Figure~\ref{fig:RFI}) are 
heavily affected by RFI across most hour angles, while spectral window 3 and 11--12 
are affected for some hour angles.}
\label{fig:RFI2}
\end{figure*}

An anomalous effect of RFI on the amplitude gain values, called ``gain compression'', was seen in the gain calibrator observations by \citep{mooley2016}.
This anomaly, where the gain response is non-linear and causes the amplitude gain values to decrease (by up to 30\%) with respect to their true values, affects pointings 
that are very close to the Clarke belt (mostly declinations between $-3$\deg and $-10$\deg, but can also affect declinations up to $+8$).
For the CNSS survey, we used gain calibrators far from the Clarke belt in order to avoid this gain compression issue.

To investigate RFI at different hour angles (HAs), we observed 10 hours in HA at a declination of $-4.5$\deg (near the satellite belt) with OTFM. 
Figure~\ref{fig:RFI2} shows the corresponding dynamic spectrum. 
Although there are some regions that seem relatively quiet for most of the HA range, more than half of the S-band is affected by severe RFI for a wide range of HAs. 
Most, but not all, of the resulting gain compression can be corrected using the switched power, since during gain compression the inverse gain amplitude is proportional to the switched power \citep[see also][]{mooley2016}.

\section{OTFM Testing Results}\label{sec:test}


\begin{figure*}
\centering
\includegraphics[width=\textwidth]{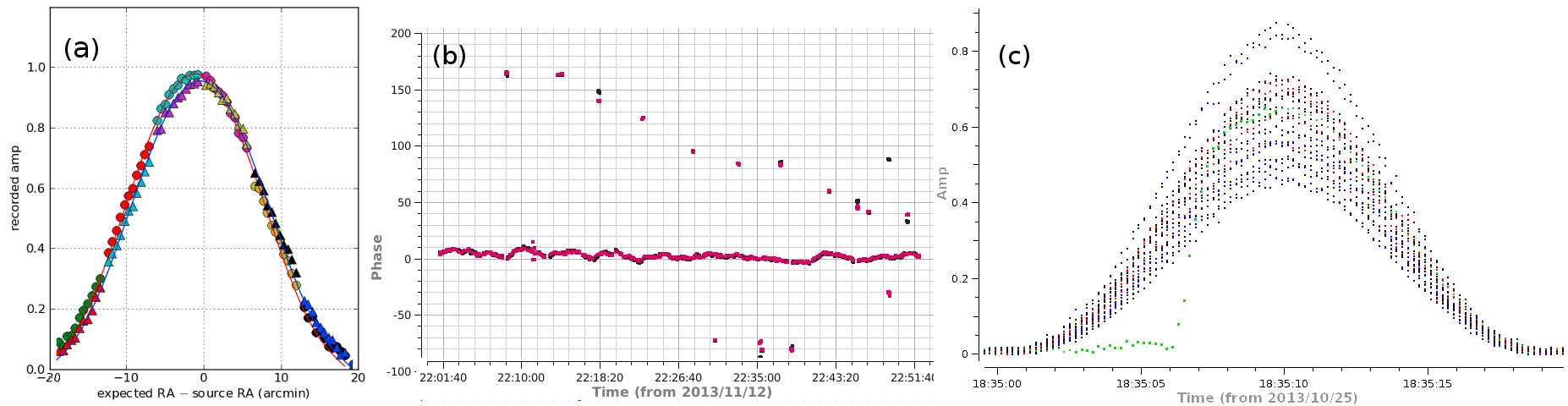}
\caption
{Some of the technical issues uncovered in OTFM data during test observations. 
We observed a bright phase calibrator source, J1229+0203 (3C273), with several back-and-forth OTFM scans.
After calibrating the phases, we took the visibility data for each antenna, vector averaged all baselines and inspected for amplitude vs. time and phase vs. time plots for RR and LL correlations. 
(a) There was a $\sim$0.6\arcmin offset (this is the average for the RR and LL beam crossings, given beam squint) between where the expected phase centers and measured phase centers. 
The source crosses the peak of the beam earlier than expected.
(b) Systematic phase errors on a fraction of the dumps on one antenna (calibrated phases deviate from zero).
(c) Some antennas were lagging behind while others were already observing the source.
In this panel, antenna ea27 (data in green) arrives at the source several seconds after the other antennas have already started the OTFM scan on the source.
Appropriate online flags were set to solve this issue. See \S\ref{sec:test} for details.}
\label{fig:tests}
\end{figure*}

In 2013, the VLA OTFM observing mode was offered in the Resident Shared Risk Observing (RSRO) program.
For the CNSS we ran some tests in this mode as part of the RSRO program and subsequently, in 2016, some additional test data were taken for the VLA Sky Survey (VLASS).
In the context of the primary science case for the CNSS, radio transient pheomena, here we give a brief description of the technical problems that were uncovered with OTFM observations.
Some of these were solved and some still exist in the data.

The test observations were undertaken as part of VLA programs TRSR0015 and TSKY0001.
We observed a bright phase calibrator source, J1229+0203 (3C273), with several back-and-forth OTFM scans.
For most of the tests, we used dump time of 0.5 s at S band and correlator phase center change every 6 arcmin (i.e. 15 dumps per primary beam, for scan rate of 1\arcmin/sec, and 12 dumps per phase center step).

\subsection{Data-recording and source-tracking issues}\label{sec:test:recording_issues}
After calibrating the phases in the test data, we took the visibility data for each antenna, vector averaged all baselines and inspected for amplitude vs. time and phase vs. time plots for RR and LL correlations (in CASA {\tt plotms}). 
We recovered several issues in the data.
1) There were systematic phase errors on a fraction of the dumps on all antennas, such that the calibrated phases deviated from zero (Figure~\ref{fig:tests} panel (b)).
This issue was traced back to the delay polynomials set at the VLA, and was fixed.
2) We found that some of the dumps at sporadic intervals were dropped (missing from the recorded data).
This problem was traced to the correlator back-end (CBE), and fixed.
3) There was a 0.6 second offset, equivalent to $\sim$0.6\arcmin~in RA (this is the average for the RR and LL beam crossings, given beam squint), between the expected phase centers and measured phase centers. 
This is shown in Figure~\ref{fig:tests} panel (a), where the source crosses the peak of the beam earlier than expected.
This issue, occurring due to the antenna position not being recorded at the appropriate timestamp, was not solved until 2016, but it alters only the timestamps in the CNSS data and does not affect the data quality in any way. 
4) We also found that some antennas were lagging behind while others were already observing the target.
This is shown in Figure~\ref{fig:tests} panel (c), where the antenna ea27 arrives at the source several seconds after the other antennas have already started the OTFM scan on the source.
Appropriate online flags were set to solve this issue.
5) In testing for VLASS, we identified an issue where the antenna trajectories did not cross the phase center position at the midpoint of the (sub)scan but were offset, 
increasing the magnitude of the smearing error given by the offset from the phase center. This was not corrected until the first observing cycle of the VLASS (in 2017).\
6) At the beginning of each OTF scan/stripe, the antennas were not yet up to tracking speed in spite of the backup scan used. 
As a consequence, the first few integrations in a row are likely to be flagged on most antennas.
7) In the VLASS test observations, ghost images of real sources were found in the image plane. 
This occurred due to correlator phase centers being incorrectly written into the dataset.
This problem was fixed only in 2016.
Although there may have been some extreme cases in the CNSS where this problem occurred, we expect only a few (or less) of the CNSS observations to be affected.

%
%
\subsection{Flux density and source position issues recovered by comparing OTFM mode data with standard pointing mode data (image plane)}\label{sec:test:OTFM_vs_PTG}

We compared the image mosaics and source catalogs from the CNSS survey, observed in OTFM mode, and the 50 deg$^2$ CNSS Pilot survey \citep{mooley2016}, observed in standard pointing (PTG) mode.
The necessary details of the observations, data processing, imaging and source cataloging for both of these surveys are given in \cite{mooley2016}.

\begin{figure}[htp]
\centering
\includegraphics[width=3.5in]{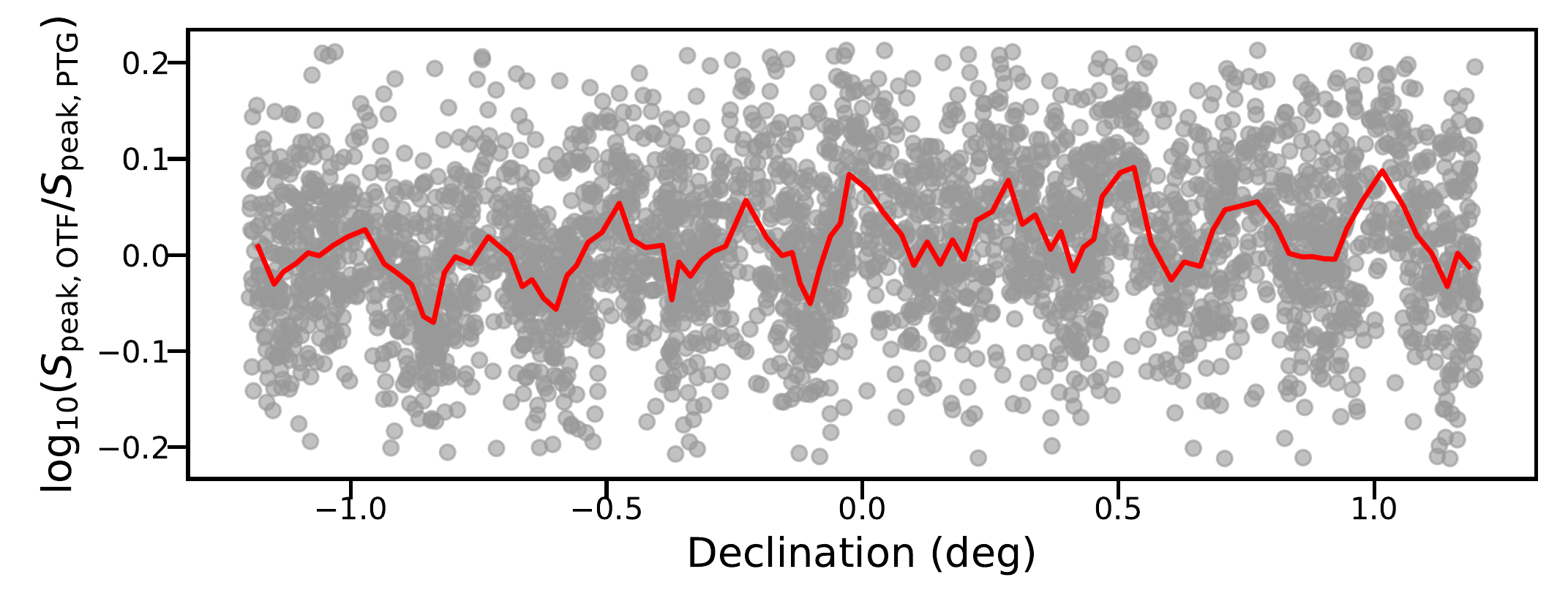}
\includegraphics[width=3.5in]{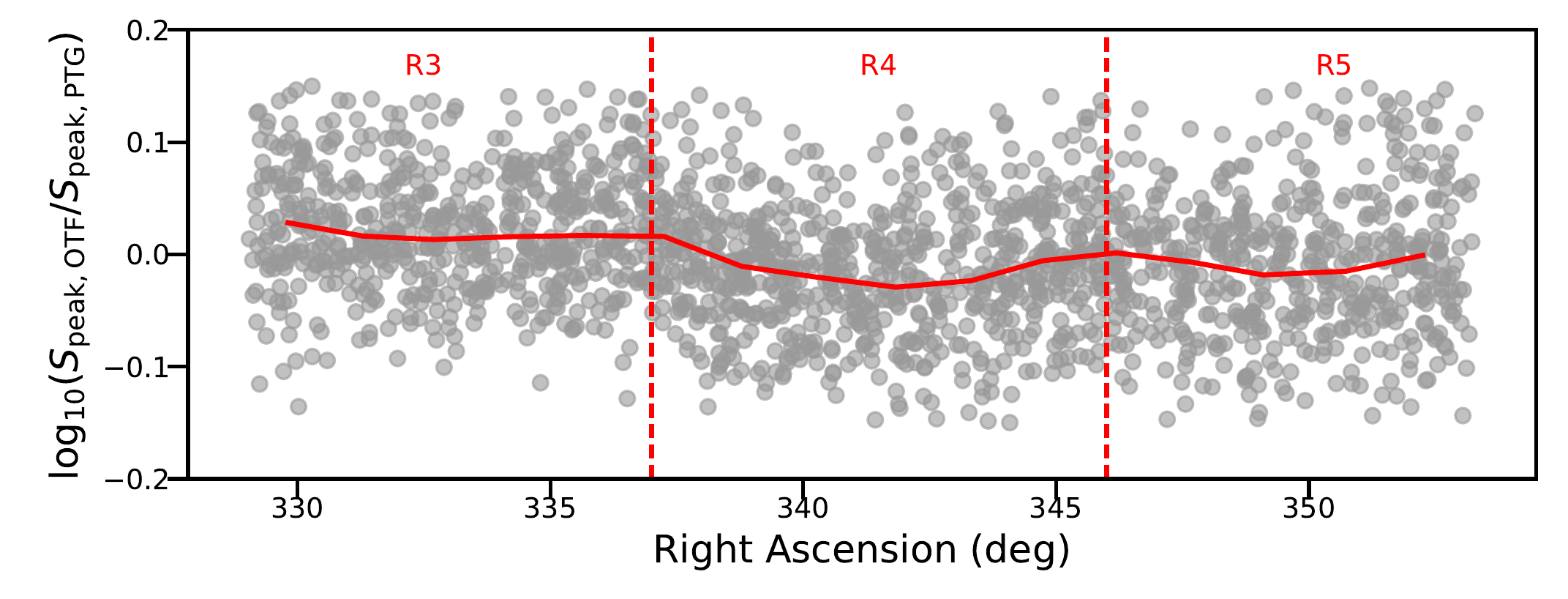}
\caption{The ratios of flux densities of point-like sources between the CNSS (OTF mode) and the CNSS Pilot survey (PTG mode), plotted as a function of right ascension (top panel) and declination (bottom panel). 
The dashed red lines in the bottom panel demarcate the different regions (R1--R12) within Stripe 82 that were observed as individual observing blocks in the CNSS.
The solid red curves represent a moving average (plotted independently for each region in the bottom panel).
The top panel shows a sinusoidal variation in the flux density ratios with declination. 
This is likely due to the primary beam corrections: narrower primary beam of the upgraded VLA with respect to the old VLA, and the smeared primary beam.
The moving average from the top panel was used to correct the flux density ratios before plotting the bottom panel.
The bottom panel suggests that there are regions that have slightly discrepant offsets in the flux densities.
See \S\ref{sec:test:OTFM_vs_PTG} for details.
}
\label{fig:S_OTF_vs_PTG}
\end{figure}

\begin{figure}[htp]
\centering
\includegraphics[width=3.5in]{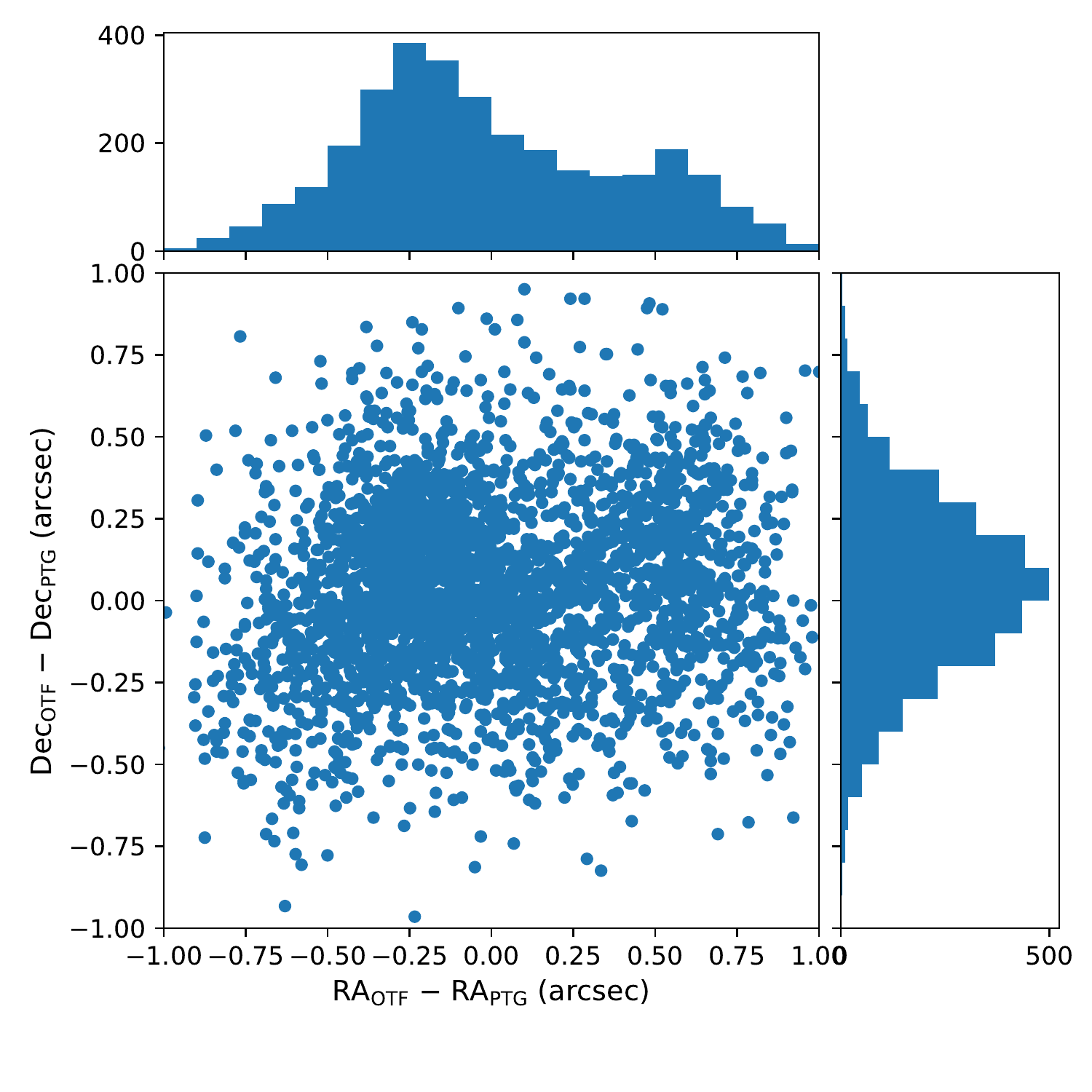}
\caption{Scatter plot of the positional offsets of point-like sources in RA and Dec between CNSS (OTFM mode) and CNSS Pilot (PTG mode) observations.
The panel to the top and panel to the right are histograms of the RA offset and Dec offset respectively.
The discrepancy is primarily in RA, and is different for different regions within Stripe 82, as shown in Figure~\ref{fig:PosOffset_OTF_PTG}.
See \S\ref{sec:test:OTFM_vs_PTG} for details.
}
\label{fig:PosOffset_ScatterHist}
\end{figure}

The ratios of flux densities of point-like sources\footnote{We selected sources having detection SNR$>7$ and the ratio of integrated-to-peak flux density 
smaller than 1.5, cf. \cite{mooley2016}} between the CNSS (center frequency 3.0 GHz) and the CNSS Pilot survey (center frequency 2.8 GHz) 
are plotted as a function of right ascension and declination in Figure~\ref{fig:S_OTF_vs_PTG}.
The ratio has a sinusoidal pattern, varying with declination, having an amplitude of $\leq$12\% (i.e. $\leq$10$^{0.05}$).
The declination values of the crests and troughs of the pattern are consistent approximately with the half power points of the primary beam and the pointing centers respectively, of the 
pointing locations of the CNSS Pilot survey \citep[see Figure 1 of ][]{mooley2016}.
This suggests that the pattern is arising largely (if not entirely) from the primary beam corrections applied to the OTFM, where we have used the VLA primary beam instead of the OTFM smeared beam, and PTG, 
where the old VLA primary beam was used instead of the narrower primary beam of the upgraded VLA \cite[cf. ][]{perley2016}.
Once the sinusoidal pattern is corrected using the moving average (solid red curve in the top panel of Figure~\ref{fig:S_OTF_vs_PTG}), the ratios of flux densities show a slight systematic 
effect ($\leqslant$6\%, i.e. 10$^{0.025}$) as a function of right ascension.
This is shown in the bottom panel of Figure~\ref{fig:S_OTF_vs_PTG}.
We are currently investigating in more detail the cause of these offsets.

\begin{figure*}[htp]
\centering
\includegraphics[width=7in]{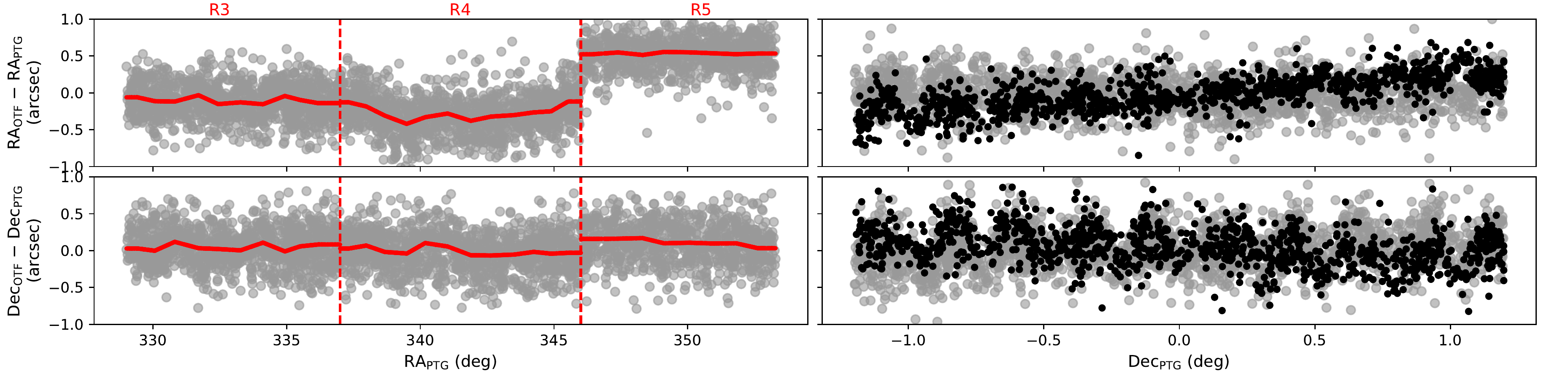}
\caption{The positional offsets of sources detected in the CNSS (OTFM mode) versus those in the CNSS Pilot survey (PTG mode), plotted as a function of RA (left panels) and Dec (right panels).
The top panels show the plots for offsets in RA and the bottom panels show the offsets in Dec.
The dashed red lines in the left panels demarcate the different regions (R1--R12) within Stripe 82 that were observed as individual observing blocks in the CNSS.
The solid red curves represent a moving average (plotted independently for each region in the bottom panel).
The black points in the right panel represent sources from a specimen region, in this case R3.
The plots indicate that there are discrepant offsets in the different regions as well as in declination.
See \S\ref{sec:test:OTFM_vs_PTG} for details.
}
\label{fig:PosOffset_OTF_PTG}
\end{figure*}

\begin{figure*}[htp]
\centering
\includegraphics[width=7in]{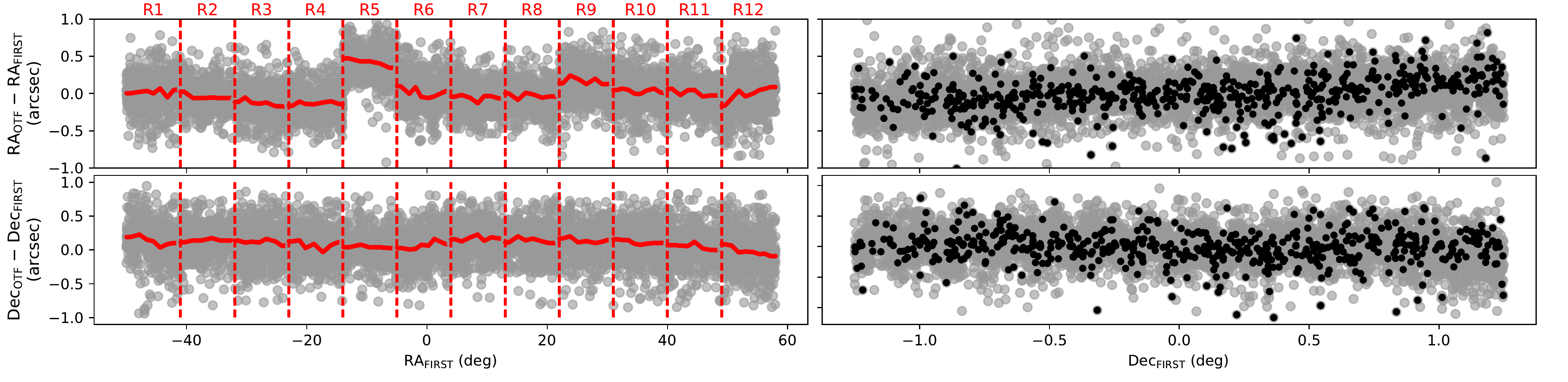}
\caption{Same as Figure~\ref{fig:PosOffset_OTF_PTG}, but comparing the source positions in the full CNSS survey (OTFM mode) with those in the FIRST survey.
See \S\ref{sec:test:OTFM_vs_PTG} for details.
}
\label{fig:PosOffset_FIRST}
\end{figure*}

The positional offsets\footnote{We note that, since we have used sources having SNR$>7$, the $\sim$3\arcsec~synthesized beam implies an uncertainty in RA and Dec of $\sim$0.2\arcsec~or better.} 
between the OTFM mode and PTG mode observations are shown in Figure~\ref{fig:PosOffset_ScatterHist}.
While no significant discrepancy is seen in Dec, there appears to be systematic uncertainties in the RA by 0.2\arcsec--0.6\arcsec.
The histogram of the RA offset is bimodal, with peaks at $-$0.25\arcsec~and 0.55\arcsec, and this suggests that the RA offset may be a function of RA and/or Dec.
In Figure~\ref{fig:PosOffset_OTF_PTG} we show the positional offsets of sources plotted as functions of RA and Dec.
The plots indicate that there are discrepant offsets, between 0\arcsec--0.5\arcsec, in the different regions as well as in declination. 
R5 is the most severely affected region.
After the positional offsets were corrected for the discrepancies as a function right ascension (left panels of Figure~\ref{fig:PosOffset_OTF_PTG}; using the moving average shown as red curves in the figure), 
a sinusoidal pattern in the offset plotted as a function of declination can be seen (right panels of Figure~\ref{fig:PosOffset_OTF_PTG}), similar to but less distinct than in the case of the ratio of 
flux densities (see above).
The RA offset in region R3 (shown as black points in the top right panel of Figure~\ref{fig:PosOffset_OTF_PTG}) also shows a monotonic increase as a function of the declination.
The CNSS Pilot survey covered only 50 deg$^2$, so in order to investigate the positional offsets over the entire Stripe 82 region we cross-matched our CNSS catalog of point-like sources with the FIRST catalog \citep[version 12feb16;][]{white1997}.
The corresponding positional offsets are shown in Figure~\ref{fig:PosOffset_FIRST}.
A similar trend as recovered with the OTFM versus PTG comparison is seen in this figure. 
R5 remains the most severely affected region, with offsets of $\sim$0.5\arcsec.
We explore the possible causes of the positional offsets in the next subsection.

\begin{figure}
\centering
\includegraphics[width=3.4in,trim={0.6cm 0 0.5cm 0.5cm},clip]{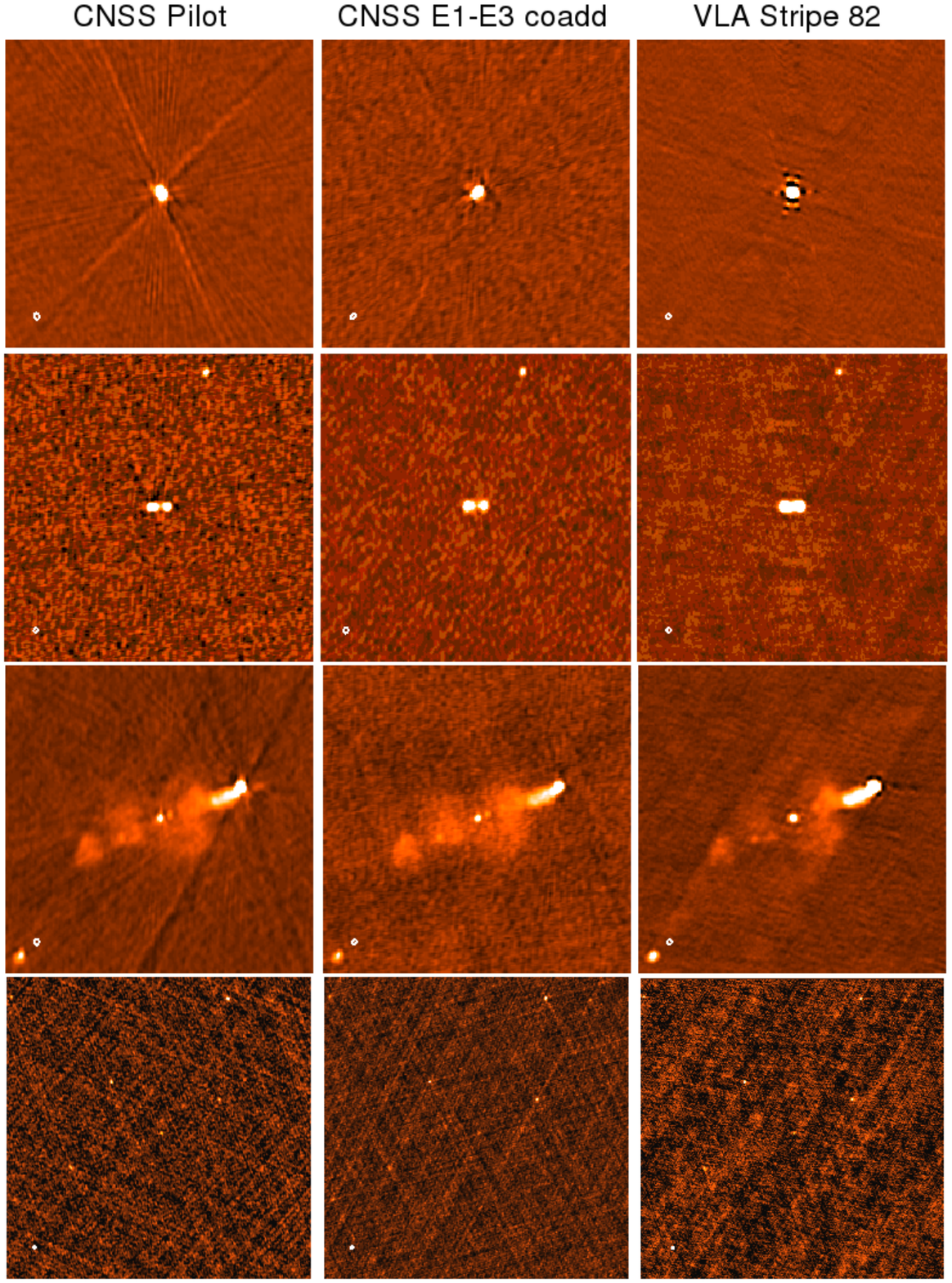}
\caption
{Image cutouts from the OTFM (middle panels) and the PTG (left and right panels) mode observations agree very well. 
From left to right, the panels show cutouts from the CNSS Pilot \citep[2.8 GHz;][]{mooley2016}, the CNSS (3.0 GHz; coaddition of epochs E1, E2, E3), and 
the VLA Stripe 82 \citep[1.4 GHz;][]{hodge2011} surveys. 
This figure shows (from top to bottom) the comparisons for a) a bright source CNSS J221947.2$-$005132 (300 mJy peak flux density, slightly resolved), 
b) a double-lobe source CNSS J221155.6$+$000447 (15--20 mJy peak flux density, resolved),
c) an extended source CNSS J221830.3$+$001219 and d) faint point-like sources (sub-mJy, cutouts centred on coordinates 22 16 27 $-$00 54 00).
The color palette used is hot metal (white represents maximum flux density, orange is intermediate and black is minimum), and in of the above cases the colorbar spans 
$-1$\,--\,$+4$ mJy (white--black), $-0.35$\,--\,$+1.50$ mJy, $-0.7$\,--\,$+3.0$ mJy and $-50$\,--\,$+500$ $\mu$Jy.
Each cutout is $2\arcmin\times2\arcmin$ except for those in the bottom panels, which are $6\arcmin\times6\arcmin$.
See \S\ref{sec:test:OTFM_vs_PTG} for details.}
\label{fig:cutouts}
\end{figure}

In regions where the discrepancies in the source flux densities and positional offsets is minimal, a comparison of single scan/sub-scan images from the CNSS and the 
CNSS Pilot surveys show excellent agreement.
Figure~\ref{fig:cutouts} shows the comparisons between the image cutouts for a) a bright point source, b) a double-lobe source, c) an extended source 
and d) faint point-like sources from the CNSS Pilot \citep[2.8 GHz;][]{mooley2016}, the CNSS (3.0 GHz; coaddition of epochs E1, E2, E3), and 
the VLA Stripe 82 \citep[1.4 GHz;][]{hodge2011} surveys.

\subsection{Comparing OTFM mode data between different epochs}\label{sec:test:OTFM_Epoch}

We also compared the single-epoch source catalogs from the CNSS survey with each other to investigate how the OTFM data compare between the different epochs.
The necessary details of the observations, data processing, imaging and source cataloging are given in \cite{mooley2016}.

\begin{figure}[htp]
\vspace{0.2cm}
\centering
\includegraphics[width=3.5in]{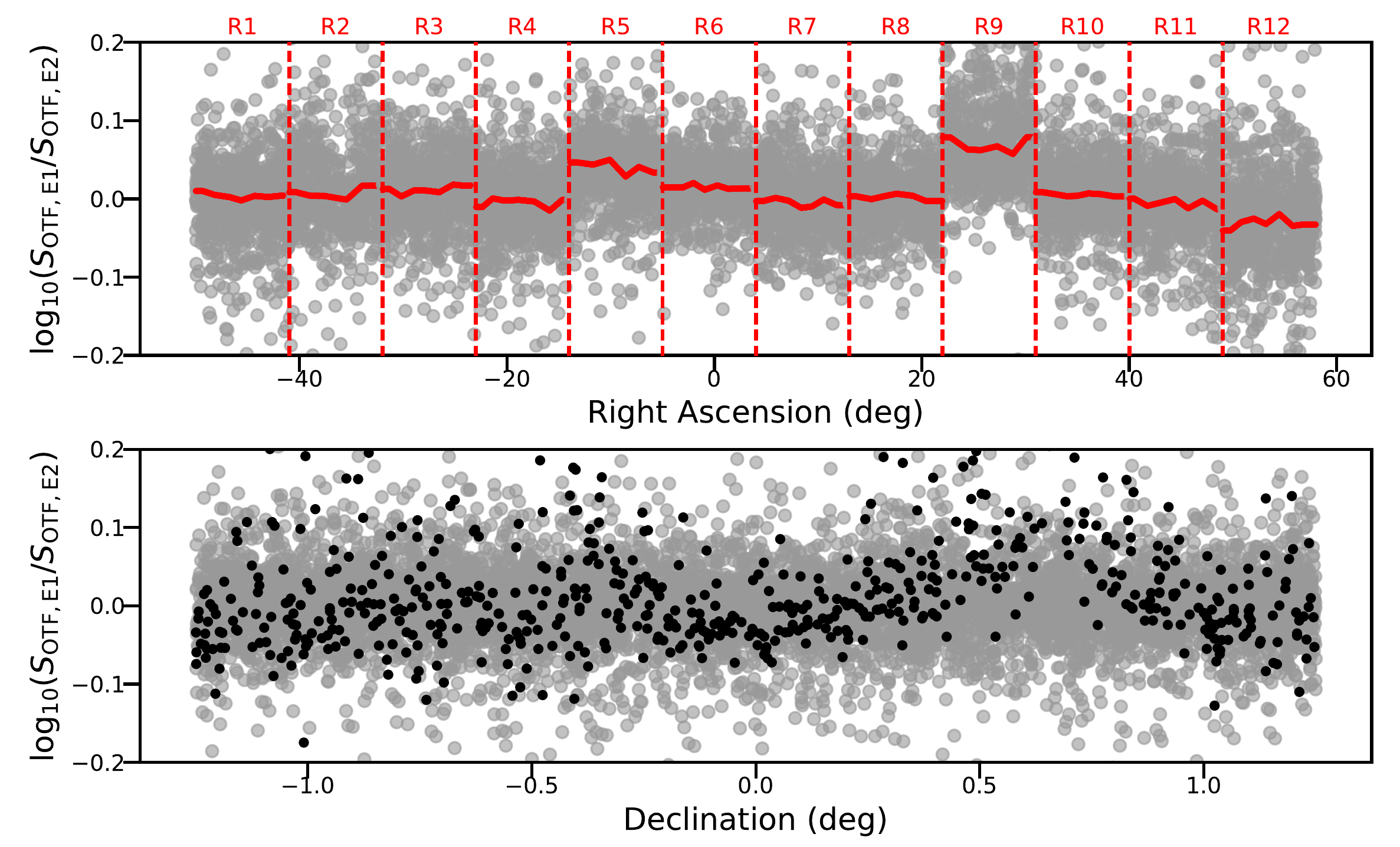}
\caption{The ratios of flux densities of point-like sources between the first two epochs (E1 and E2; grey points) of the CNSS, plotted as a function right ascension (top panel) and declination (bottom panel). 
The dashed red lines demarcate the different regions (R1--R12) within Stripe 82 that were observed as individual observing blocks in the CNSS.
The top panel shows systematic offset between the flux densities in some of the regions (e.g. R9).
The solid red curves represent a moving average, plotted independently for each region.
The moving average from the top panel was used to correct the flux density ratios before plotting the bottom panel.
The black points in the bottom panel represent sources from a specimen region, in this case R9.
The flux density offset may be due to RFI-induced gain compression; this is currently under investigation.
See \S\ref{sec:test:OTFM_Epoch} for details.
}
\label{fig:S_OTF_Epoch}
\end{figure}

\begin{figure}[htp]
\centering
\includegraphics[width=3.5in]{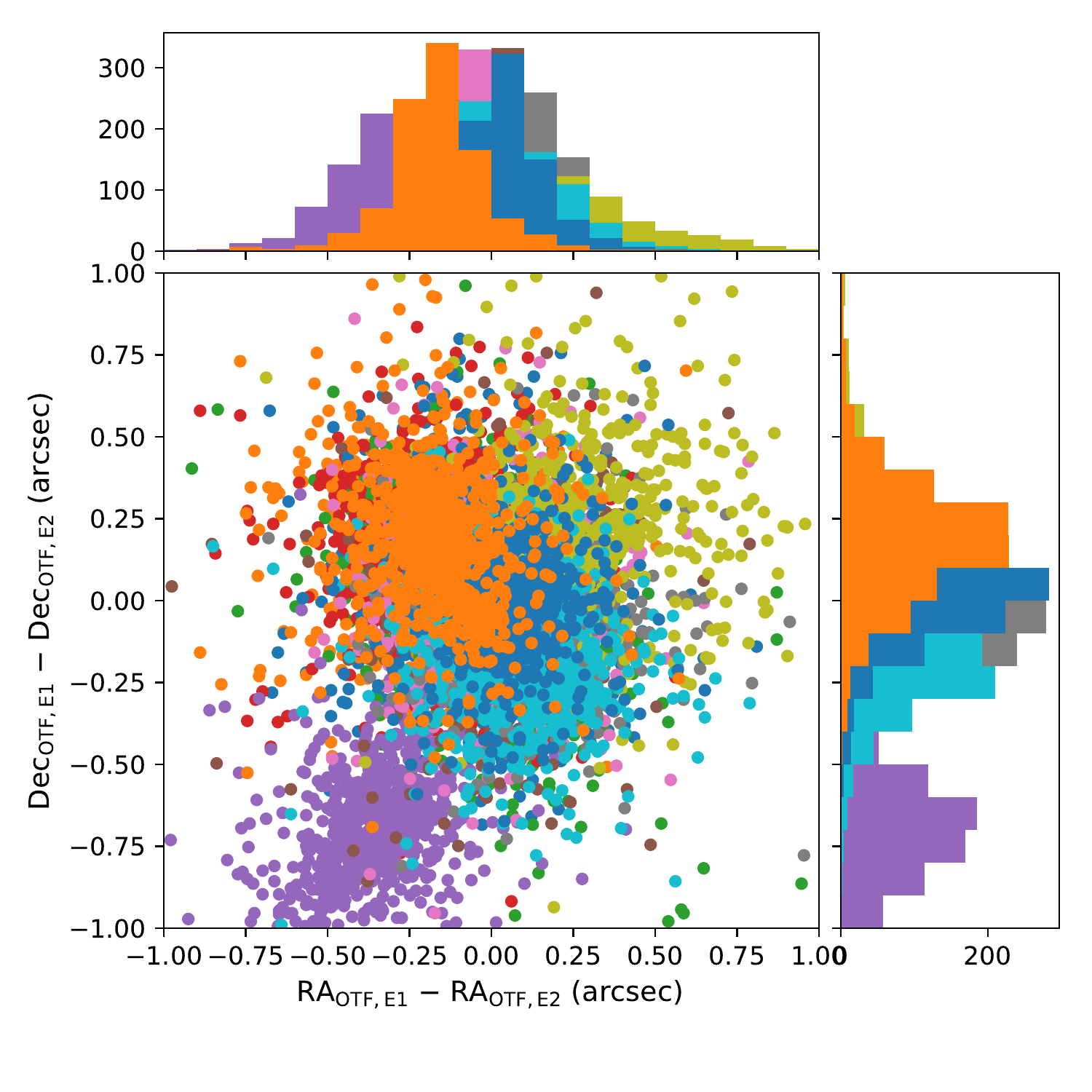}
\includegraphics[width=3in]{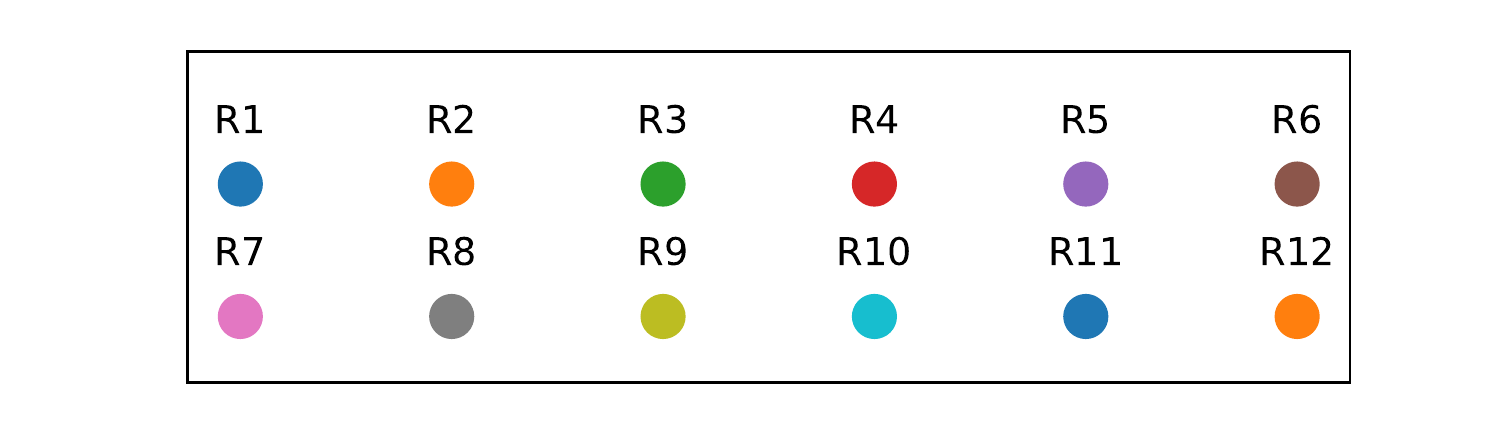}
\caption{
Scatter plot of the positional offsets of point-like sources in RA and Dec between CNSS epoch 1 and epoch 2 (observed in OTFM mode).
The different regions (R1--R12) of CNSS, observed in different scheduling blocks with the VLA, are color-coded (see legend below).
The panel to the top and panel to the right are histograms of the RA offset and Dec offset respectively.
There are systematic offsets of $\leqslant$0.25\arcsec~in RA and Dec for most of the regions.
For region R5 the offsets are severe, $-$0.4\arcsec and $-$0.7\arcsec~ in RA and Dec are respectively.
The positional offsets are likely due to inaccurate coordinates used for the phase calibrator pointings. 
See \S\ref{sec:test:OTFM_Epoch} for details.
}
\label{fig:PosOffset_OTF_Epoch}
\end{figure}

The ratios of flux densities of point-like sources between the epochs E1 and E2 of the CNSS are plotted as a function right ascension and 
declination in Figure~\ref{fig:S_OTF_Epoch}.
the ratios of flux densities show significant systematic offset from unity (0\%--26\%, i.e. $\leqslant$10$^{0.1}$) as a function of right ascension.
The offset for each region, R1--R12, appears to be different (and independent any other region; as with the OTFM versus PTG case described in the previous subsection), with R9 being the most extreme case.
We corrected the flux density ratios using multiplicative factors found with a moving average (solid red curve in the upper panel of Figure~\ref{fig:S_OTF_Epoch}), 
and plotted the corrected ratios as a function of declination (upper panel of Figure~\ref{fig:S_OTF_Epoch}).
The ratios for the different regions are generally within 12\% (i.e. $\leqslant$10$^{0.05}$), but for regions R9 there is a large discrepancy, $\sim$25\%, at Dec$\simeq$0.6 deg
(black points in the bottom panel of Figure~\ref{fig:S_OTF_Epoch}.
Manual inspection of the Dec$\simeq$0.6 deg region reveals that epoch E2 image is noisy and has an slightly elongated beam.
The flux density offset may be due to RFI-induced gain compression.
This issue will be investigated further in CNSS Paper III (Mooley et al. in prep).

The positional offsets between the OTFM epochs E1 and E2 are shown in Figure~\ref{fig:PosOffset_OTF_Epoch}, where the different regions are color-coded.
Although there are systematic offsets for most of the regions, the median offsets in RA and Dec are $\leqslant$0.25\arcsec. 
Region R5 is an exception, where the offsets in RA and Dec are $-$0.4\arcsec and $-$0.7\arcsec~ respectively.
It is likely that the positional offsets are caused by incorrect positions for some of the phase calibrator sources.
For example, in the observing for region R5, the VLA calibrator J2323$-$0317 was used in epoch E1, while a bright source (quasar) from the VLA Stripe 82 survey \citep{hodge2011}, J224730+000006, 
was used in epoch E2 (since J2323$-$0317 showed some indication for gain compression in E1).
While J2323$-$0317 has an accurate source position, we have found that in the image plane J224730+000006 is offset by $\sim$0.8 arcsec with respect to its position reported by \citep{hodge2011}.
This is consistent with the offset seen in Figure~\ref{fig:PosOffset_ScatterHist}.
The epoch-to-epoch positional offsets in R5 also explains the severe offsets seen in the OTFM versus PTG comparisons (\S\ref{sec:test:OTFM_vs_PTG}, 
Figures~\ref{fig:PosOffset_OTF_PTG}--\ref{fig:PosOffset_FIRST}).


\section{Summary \& Discussion}\label{sec:summary}


We have developed and tested a new imaging capability for the VLA in which the antennas are driven at a non-sidereal rate and visibilities are recorded continuously. We find that OTFM observations significantly reduce the slew-and-settle overheads as compared with pointed observations requiring on-source time of less than $\sim$30 s per pointing. Through the use of newly-developed imaging techniques in the CASA package, we have also demonstrated that the flux densities of sources can be reliably and accurately reproduced not only between two epochs observed using OTFM, but also between pointed observations and those observed with the OTFM technique.

The implementation of OTFM substantially improves the efficiency of conducting wide-field shallow surveys with the VLA. The VLA's survey capabilities can be compared with Square Kilometer Array pathfinders that are undergoing commissioning and will soon start science operations. As noted in \S\ref{sec:intro}, this includes the Australian Square Kilometre Array Pathfinder \citep[ASKAP;][]{johnston2008}, the Apertif instrument on the Westerbork Synthesis Radio Telescope \citep{aab+18,rab+11}, and MeerKAT \citep{sbc+16,bve+16}. The proper figure of merit for comparing different imaging interferometers is the survey speed (SS; \S\ref{sec:intro:SS}), expressed as:\\
${\rm SS} \propto {\rm BW} \times \Omega({\rm A_e/T_{sys}})^2$ 
where BW is the bandwidth, $\Omega$ is the field-of-view, A$_e$ is the total collecting area times the aperture efficiency and T$_{sys}$ is the antenna system temperature \citep{cordes2008}. 
For the VLA L band and S band we estimate\footnote{For details, see~ \url{https://science.nrao.edu/facilities/vla/docs/manuals/obsguide/modes/mosaicking} and \cite{mooley2013}.} the survey speeds to be about 15.3~deg$^2$\,hr$^{-1}$ and 13.4~deg$^2$\,hr$^{-1}$ respectively.
It is OTFM that makes the VLA a competitive survey instrument, while ASKAP and Apertif achieve high SS values using 
phased array feeds, and MeerKAT employs large numbers of small diameter antennas having high $A_e/T_{sys}$. 

OTFM opens up several new applications. In this work we have used OTFM for the Caltech-NRAO Stripe 82 Survey (CNSS) which was carried out primarily to search for long-duration transients\footnote{We define slow or long-duration transients in accordance with literature as those having timescales $>$1s. See \url{http://tauceti.caltech.edu/kunal/radio-transient-surveys.html} for an up-to-date list of radio surveys aimed at exploring slow transient phenomena.}.  The large fields-of-view that are required are driven primarily by the rarity of radio transients \citep[e.g][]{mooley2016,mkc+17}. Likewise, recent near-real-time surveys \citep{ofek2011,mooley2016} have stressed the unique advantage of associating radio sources with optical counterparts. This tends to favor multi-epoch shallow surveys which bring the observable transient population closer in distance, improving the ability to find optical/infrared counterparts and to characterize host galaxies and/or progenitors. Radio transient searches therefore call for wide-field shallow surveys.

OTFM has also been applied in an all-sky continuum survey and in the search for the electromagnetic counterparts of gravitational waves (GW). 
The VLA Sky Survey (VLASS) is using OTFM to efficiently survey the entire sky in the continuum band between 2 and 4 GHz above a declination of $-40^\circ$ with a resolution of 2.5$^{\prime\prime}$ \citep{mbc+17}. 
In the recent binary neutron star merger event GW\,170817, radio observations played a key role in the study of the afterglow component \citep{hcm+17,mnh+18,max+18}. 
However, looking ahead to the near future, while the sensitivity of existing GW detectors will be improved, the median sky localizations will remain $\sim$150 deg$^2$ \citep{aaa+16}. 
These large, elongated error regions lend themselves well to OTFM. 
In \citet{mfm+18}, we carried out a successful proof-of-concept demonstration toward the binary black hole merger event GW\,151226, using OTFM to image a 100 deg$^2$ region over three epochs 
separated by 1.5 and 6 months post-merger. 
OTFM of the entire error region will not likely be as efficient as optical/UV searches for the thermal kilonova component, nor will they be as optimal as using 
volume-limited galaxy searches \citep[e.g.][]{sch+16} but it occupies a valuable niche. 
As noted in \citet{mfm+18}, there will be GW events with solar/lunar constraints, events that occur in obscured environments, incomplete galaxy catalogs, and faint or 
non-existent kilonova signals \citep[e.g. NS-NS mergers with large binary mass;][]{metzger2017}
In these cases blind radio searches of the GW localization, aided by valuable ``before" images from the VLASS (or OTFM observations immediately after the merger), 
may be our best hope at detecting the afterglow signature of a GW event.


\acknowledgments{\it The authors would like to thank Urvashi Rau and Sanjay Bhatnagar for developing (and implementing in CASA) the algorithms used extensively in this work. 
We are also grateful to Barry Clark, Michael Rupen and other NRAO staff who provided extensive support during the commissioning of the OTFM observing mode. 
The National Radio Astronomy Observatory is a facility of the National Science Foundation operated under cooperative agreement by Associated Universities, Inc. 
K.P.M. is currently a Jansky Fellow of the National Radio Astronomy Observatory.}

\appendix

{\tt OTFSim} is similar to {\tt SCHED} used for VLBA, in some respects. The algorithm for the simulator script is relatively straightforward:
\begin{enumerate}
 \item Start the antennas at an arbitrary azimuth (AZ) and elevation (EL). Count the time and for each time increment check the user-specified order of scans for the following.
 \item If a calibrator is to be observed next, then slew to the current AZ and EL of that calibrator source and track the source for 
 the user-specified duration.
 \item If OTFM observations to be carried out next, then 
 slew to the required part of the survey region, then slew the antennas across a single stripe in right ascension along constant declination with a user-specified slew speed. 
 Repeat the slew in adjacent stripes till the user-specified duration is complete.
\end{enumerate}


\end{document}